\documentclass[fleqn,usenatbib]{mnras}

\usepackage[dvipdfmx]{graphicx,color}	
\usepackage{graphicx}	
\usepackage{amsmath}	
\usepackage{amssymb}	

\usepackage{footnote}
\usepackage{xspace}

\newcommand{\hGpc}{{\ifmmode{h^{-1}{\rm Gpc}}\else{$h^{-1}$Gpc}\fi}}
\newcommand{\hMpc}{{\ifmmode{h^{-1}{\rm Mpc}}\else{$h^{-1}$Mpc}\fi}}
\newcommand{\hkpc}{{\ifmmode{h^{-1}{\rm kpc}}\else{$h^{-1}$kpc}\fi}}
\newcommand{\invGyr}{{\ifmmode{{\rm Gyr}^{-1}}\else{${\rm Gyr}^{-1}$}\fi}}
\newcommand{\kms}{{\ifmmode{{\rm km~s}^{-1}}\else{${\rm km~s}^{-1}$}\fi}}
\newcommand{\invHz}{{\ifmmode{{\rm Hz}^{-1}}\else{${\rm Hz}^{-1}$}\fi}}

\newcommand{\hMsun}{{\ifmmode{h^{-1}{\rm {M_{\odot}}}}\else{$h^{-1}{\rm{M_{\odot}}}$}\fi}}
\newcommand{\Mstar}{{\ifmmode{M_{*}}\else{$M_{*}$}\fi}}

\usepackage{breakurl}

\definecolor{mygray}{gray}{0.7}

\usepackage[dvipsnames]{xcolor}
\usepackage[normalem]{ulem} 

\title[Cosmic variance forecasts of AGN in Uchuu--$\nu^2$GC]{Uchuu--$\nu^2$GC galaxies and AGN: Cosmic variance forecasts of high-redshift AGN for {\it JWST}, {\it Euclid}, and LSST}

\author[T. Oogi et al.]{
Taira Oogi,$^{1,2}$\thanks{E-mail: oogi@cosmos.phys.sci.ehime-u.ac.jp (TO)}
Tomoaki Ishiyama,$^{2}$ Francisco Prada,$^{3}$ Manodeep Sinha,$^{4,5}$ 
\newauthor{ Darren Croton,$^{4,5}$ Sof\'{i}a A. Cora,$^{6,7}$ Eric Jullo,$^{8}$ Anatoly A. Klypin,$^{9,10}$} 
\newauthor{ Masahiro Nagashima,$^{11}$ J.~L\'opez Cacheiro,$^{12}$ Jos\'{e} Ruedas,$^{3}$}
\newauthor{ Masakazu A. R. Kobayashi,$^{13,14}$ Ryu Makiya,$^{15,16}$}
\\
$^{1}$Research Center for Space and Cosmic Evolution, Ehime University, 2-5, Bunkyo-cho, Matsuyama, Ehime 790-8577, Japan\\
$^{2}$Digital Transformation Enhancement Council, Chiba University, 1-33, Yayoi-cho, Inage-ku, Chiba 263-8522, Japan\\
$^{3}$Instituto de Astrof\'{i}sica de Andaluc\'{i}a (CSIC), Glorieta de la Astronom\'{i}a, E-18080 Granada, Spain\\
$^{4}$Centre for Astrophysics \& Supercomputing, Swinburne University of Technology, 1 Alfred St, Hawthorn, VIC 3122, Australia\\
$^{5}$ARC Centre of Excellence for All Sky Astrophysics in 3 Dimensions (ASTRO 3D), Australia\\
$^{6}$Instituto de Astrof\'isica de La Plata (CCT La Plata, CONICET,UNLP), Observatorio Astron\'omico, Paseo del Bosque,\\ B1900FWA, La Plata, Argentina\\
$^{7}$Facultad de Ciencias Astron\'omicas y Geof\'{\i}sicas, Universidad Nacional de La Plata (UNLP), Observatorio Astron\'omico,\\ Paseo del Bosque, B1900FWA La Plata, Argentina\\
$^{8}$Aix Marseille Univ, CNRS, CNES, LAM, F-13388 Marseille, France\\
$^{9}$Astronomy Department, New Mexico State University, Las Cruces, NM, USA\\
$^{10}$Department of Astronomy, University of Virginia, Charlettesville, VA, USA\\
$^{11}$Faculty of Education, Bunkyo University, 3337 Minami-Ogishima, Koshigaya-shi, Saitama 343-8511, Japan\\
$^{12}$Centro de Supercomputaci\'on de Galicia (CESGA), Avenida de Vigo, s/n Campus Sur, E-15705 Santiago de Compostela, Spain\\
$^{13}$National Institute of Technology, 701-2 Higashiasakawa-machi, Hachioji, Tokyo 193-0834, Japan\\
$^{14}$KOSEN-KMITL, 1st Chalongkrung Road, Ladkrabang, Bangkok 10520, Thailand\\
$^{15}$Institute of Astronomy and Astrophysics, Academia Sinica, Astronomy-Mathematics Building, AS/NTU, No. 1, Section 4,\\ Roosevelt Road, Taipei 10617, Taiwan\\
$^{16}$Kavli Institute for the Physics and Mathematics of the Universe, Todai Institutes for Advanced Study, the University of Tokyo,\\ Kashiwa, Japan 277-8583 (Kavli IPMU, WPI)\\
}

\date{Accepted XXX. Received YYY; in original form ZZZ}

\pubyear{2023}

\begin{document}
\label{firstpage}
\pagerange{\pageref{firstpage}--\pageref{lastpage}}

\maketitle

\begin{abstract}
Measurements of the luminosity function of active galactic nuclei (AGN) at high redshift ($z\gtrsim 6$) are expected to suffer from field-to-field variance, including cosmic and Poisson variances. Future surveys, such as those from the {\it Euclid} telescope and {\it James Webb Space Telescope (JWST)}, will also be affected by field variance.
We use the Uchuu simulation, a state-of-the-art cosmological $N$-body simulation with 2.1 trillion particles in a volume of $25.7~\mathrm{Gpc}^3$, combined with a semi-analytic galaxy and AGN formation model, 
to generate the Uchuu-$\nu^2$GC catalog, publicly available, that allows us to investigate the field-to-field variance of the luminosity function of AGN.
With this Uchuu--$\nu^2$GC model, we quantify the cosmic variance as a function of survey area, AGN luminosity, and redshift. In general, cosmic variance decreases with increasing survey area and decreasing redshift. We find that at $z\sim6-7$, the cosmic variance depends weakly on AGN luminosity.
This is because the typical mass of dark matter haloes in which AGN reside does not significantly depend on luminosity.
Due to the rarity of AGN, Poisson variance dominates the total field-to-field variance, especially for bright AGN. We also examine the effect of parameters related to galaxy formation physics on the field variance.
We discuss uncertainties present in the estimation of the faint-end of the AGN luminosity function from recent observations, and extend this to make predictions for the expected number of AGN and their variance for upcoming observations with {\it Euclid}, {\it JWST}, and the Legacy Survey of Space and Time (LSST).
\end{abstract}

\begin{keywords}
methods: numerical -- catalogues -- galaxies: formation -- galaxies: nuclei -- large-scale structure of Universe -- cosmology: theory.
\end{keywords}



\section{Introduction}

Observations of the number density of galaxies and active galactic nuclei (AGN), and their redshift dependence across cosmic time, have yielded a number of important insights into their formation processes in a cosmological context. Our current knowledge has been advanced by comparing observations and theoretical predictions derived from semi-analytic models and cosmological hydrodynamical simulations (e.g. \citealt{Somerville_Dave2015}). In particular, the AGN luminosity function (LF) is known to be shaped by various physical processes, such as triggers of gas accretion onto the black hole, the gas accretion rate, accretion disc states, the radiative efficiency, the AGN light curve, and their obscured (or observable) fraction (e.g. \citealt{Hirschmann2012}; \citealt{Fanidakis2012}; \citealt{Shirakata2019}; \citealt{Griffin2019}). All these processes can be modelled and explored in detail.

However observations from surveys are limited by cosmic variance, and one must quantify this variance to properly constrain galaxy and AGN formation models. The faint-end of the AGN LF at $z \sim 6$ has non-negligible errors on the slope and normalisation (e.g. \citealt{Willott2010}; \citealt{Matsuoka2018}). These uncertainties also affect the detectability of AGN and the design strategy for future AGN surveys and their instrument requirements; e.g. the kinds of future instruments needed, and the types of surveys targeting AGN that will better help nail down such uncertainties. 

Similarly, measurements of the galaxy ultraviolet LF across the epoch of reionisation (\citealt{Bouwens2015}; \citealt{Livermore2017}) have large uncertainties at the faint-end (\citealt{Atek2018}; \citealt{Ishigaki2018}; \citealt{Yue2018}). And while data from new surveys and facilities, such as {\it James Webb Space Telescope (JWST)} (\citealt{Gardner2006}), {\it Euclid} (\citealt{Euclid2011}), and the Legacy Survey of Space and Time (LSST) (\citealt{LSST2009}), will dramatically increase the volume of high-redshift ($z \gtrsim 6$) galaxy and AGN samples, a precise measurement of the LF will remain limited due to the inevitable finite survey volume. It is thus essential to quantify this uncertainty for current and future surveys and identify the required precision needed by them to accurately measure the relative galaxy and AGN contributions to cosmic reionisation (\citealt{Giallongo2019}; \citealt{Grazian2020}).

A number of studies have theoretically evaluated the cosmic variance of galaxies at high redshift (e.g. \citealt{Somerville2004}; \citealt{Trenti2008}; \citealt{Ogura2020}; \citealt{Bhowmick2020}; \citealt{Ucci2021}). \citet{Ogura2020} used a semi-analytic model of galaxy and AGN formation, the New Numerical Galaxy Catalog ($\nu^2$GC; \citealt{Makiya2016}; \citealt{Shirakata2019}), to evaluate the cosmic variance of H$\alpha$ emitters at $z=0.4$. They also quantified how cosmic variance depends on survey volume. \citet{Bhowmick2020} used the large and high resolution cosmological hydrodynamical simulation, \textsc{BlueTides} (\citealt{Feng2016}), to investigate the cosmic variance of $z>7$ galaxies. And \citet{Ucci2021} investigated the cosmic variance of high-$z$ galaxies during the epoch of reionisation, and how cosmic variance depends on the galaxy and AGN model parameters. Commonly, in such studies cosmic variance is evaluated using the two-point correlation function (e.g. \citealt{Somerville2004}; \citealt{Trenti2008}; \citealt{Moster2011}), advantageous for sparse populations, such as AGN, when limited to smaller volumes, as has been the case with simulations up until recently.

Theoretical work has also been done to predict the AGN population at $z\gtrsim6$ using semi-analytic models (\citealt{Ricarte2018}; \citealt{Griffin2020}; \citealt{Piana2021}) and cosmological hydrodynamical simulations (\citealt{Feng2016}; \citealt{Waters2016}; \citealt{Habouzit2019}; \citealt{Marshall2020, Marshall2021}; \citealt{Ni2020, Ni2020b} -- see also \citet{Amarantidis2019} for a compilation of predictions using such model). However, these studies have not examined the cosmic variance of the AGN LF due to their insufficient simulation box sizes.

In this paper, we explore the cosmic variance of AGN by using the semi-analytic model $\nu^2$GC (\citealt{Makiya2016}; \citealt{Shirakata2019}). Specifically, we combine this model with the Uchuu simulation, a large cosmological dark matter $N$-body simulation (\citealt{Ishiyama2021}), from which we obtain the merger trees of dark matter haloes. Uchuu has an unprecedented box size of side length $2.0~h^{-1}\,\mathrm{Gpc}$ with sufficient mass resolution to resolve the dwarf galaxy scale. This enables us to investigate various statistical properties of rare objects with low space densities, such as bright galaxies and AGN. This new Uchuu-$\nu^2$GC galaxy and AGN catalog is released to public along with this paper through the Skies and Universes website\footnote{\url{http://www.skiesanduniverses.org}}. Taking advantage of its large co-moving volume, we evaluate the cosmic variance of AGN by their direct counting, instead of using the two-point correlation function as in previous works (e.g. \citealt{Somerville2004}). We focus on AGN at $z\gtrsim6$, mimicking the redshift range covered by upcoming JWST and Euclid AGN surveys, which will significantly increase the AGN statistics.

The remainder of this paper is organized as follows. In Section 2, we outline the semi-analytic model of galaxy and AGN formation used to generate our AGN statistics. The cosmic variance predicted from this model is presented in Section 3. In Section 4, we discuss the uncertainties in these predictions, and summarize our findings.

\section{Methods}
\label{sec:methods}

\subsection{The Uchuu simulation}

A distinguishing feature of this study is the state-of-the art cosmological dark matter $N$-body simulation on which it is based, the Uchuu simulation (\citealt{Ishiyama2021}). The Uchuu simulation evolved an $12800^3$ dark matter (DM) particles within a box of side length $2.0~h^{-1}\,\mathrm{Gpc}$, resulting in a DM particle mass of $3.27 \times 10^{8}~h^{-1}\,{\rm M}_{\odot}$. The cosmological parameters adopted by Uchuu are consistent with the latest results of the Planck satellite (\citealt{Planck2020}): $\Omega_0 = 0.3089, \Omega_{\mathrm{b}} = 0.0486, \lambda_0 = 0.6911, h=0.6774, n_{\mathrm{s}} = 0.9667$, and $\sigma_8 = 0.8159$. Uchuu was performed with the code \textsc{greem} (\citealt{Ishiyama2009}; \citealt{Ishiyama2012}). A total of 50 snapshots of the simulation are stored between $z=14$ and $z=0$. DM haloes are identified in each snapshot with the \textsc{rockstar} phase-space halo finder (\citealt{Behroozi2013a}). We use haloes with more than 40 particles, which corresponds to mass larger than $1.3\times10^{10}~h^{-1}\,{\rm M}_{\odot}$. DM halo merger trees are constructed from the \textsc{rockstar} halo catalog using a modified version of the \textsc{consistent trees} code (\citealt{Behroozi2013b}). Further details of the Uchuu simulation and the post-processing steps are given in \citet{Ishiyama2021}. The Uchuu DR1 data products are publicly available at the Skies \& Universes website\footnote{\url{http://www.skiesanduniverses.org/Simulations/Uchuu/}}.

\subsection{The \texorpdfstring{$\nu^2$}{ν\^2}GC semi-analytic model}
\label{subsec:model}

$\nu^2$GC is a semi-analytic model of galaxy and AGN formation, and is an extension of the Numerical Galaxy Catalog ($\nu$GC; \citealt{Nagashima2005}). The model runs on the DM halo merger trees obtained from high resolution cosmological $N$-body simulations. Here, we use the merger trees from the Uchuu simulation described above, which allows us to assess the cosmic variance with direct object counts.

$\nu^2$GC includes key baryonic processes for galaxy formation: radiative gas cooling and disc formation in DM haloes, star formation, supernova feedback, chemical enrichment, galaxy mergers and disc instabilities, which trigger starbursts and lead to the growth of bulges and supermassive black holes (SMBHs), and AGN feedback. Further details of this galaxy formation model are given in \citet{Makiya2016} and \citet{Shirakata2019}. Here, we briefly summarise the relevant physics for the current work, notably that related to AGN (see \citealt{Shirakata2019}; \citealt{Oogi2020} for a more detailed description).

In the model, SMBHs grow via mergers and disc instabilities that trigger starbursts, during which gas moves toward the central region of the galaxy and is accreted. In this paper, we focus on the disc instability process. The radius normalised by the disc scale radius, $R_{\mathrm{gas}} / r_{\mathrm{ds}}$, within which the gas migrates to the bulge, is determined as
\begin{equation}
\frac{R_{\mathrm{gas}}}{r_{\mathrm{ds}}} = (1-f_{\mathrm{g}}) f_{\mathrm{d}} f_{\mathrm{bar}},
\end{equation}
where $f_{\mathrm{d}}$ is the disc mass fraction of a galaxy, $f_{\mathrm{g}}$ is the gas mass fraction in the disc, and $f_{\mathrm{bar}}$ is a free parameter, which controls the strength of the disc instability. This formulation is motivated by the model of the merger induced gas accretion in \citet{Hopkins2009ApJ...691.1168H}. Assuming an exponential surface density profile for the gas disc, the gas mass in the disc, which migrates to the bulge and is exhausted by a starburst, $\Delta M_{\mathrm{dg,DI}}$, is given by 
\begin{equation}
\Delta M_{\mathrm{dg,DI}} = M_{\mathrm{dg}} \times \left\{ 1 - \left( 1 + \frac{R_{\mathrm{gas}}}{r_{\mathrm{ds}}} \right) \exp(-R_{\mathrm{gas}} / r_{\mathrm{ds}}) \right\},
\end{equation}
where $M_{\mathrm{dg}}$ is the gas mass of the disc.

The accreted gas mass on to a SMBH, $\Delta M_{\mathrm{BH}}$, is given by:
\begin{equation}
\Delta M_{\mathrm{BH}} = f_{\mathrm{BH}} \Delta M_{\mathrm{star,burst}},
\end{equation}
where $\Delta M_{\mathrm{star,burst}}$ is the stellar mass formed throughout a starburst. The parameter $f_{\mathrm{BH}}$ is set to $0.02$ to match the observed correlation between the masses of host bulges and their SMBHs at $z=0$ (\citealt{McConnell2013}). The gas accretion rate is described by:
\begin{equation}
\dot{M}_{\mathrm{BH}} (t) = \frac{\Delta M_{\mathrm{BH}}}{t_{\mathrm{acc}}}\exp \left(- \frac{t-t_{\mathrm{start}}}{t_{\mathrm{acc}}}\right),
\end{equation}
where $t_{\mathrm{acc}}$ is the accretion time-scale and $t_{\mathrm{start}}$ is the start time of the accretion.

We assume that $t_{\mathrm{acc}}$ is determined by the loss of angular momentum of the accreted gas in the nuclear region, in addition to the dynamical time-scale of the host galaxy's bulge, $t_{\mathrm{dyn, bulge}}$:
\begin{equation}
t_{\mathrm{acc}} = \alpha_{\mathrm{bulge}} t_{\mathrm{dyn, bulge}} + t_{\mathrm{loss}},
\end{equation}
where $\alpha_{\mathrm{bulge}}$ is a free parameter, and $t_{\mathrm{loss}}$ is the time-scale for angular momentum loss \citep{Shirakata2019}. The parameter $\alpha_{\mathrm{bulge}}$ is set to $0.58$ to ensure that the bright end of the AGN LF matches the observations (\citealt{Ueda2014}; \citealt{Aird2015}). In other words, the accretion rate of bright AGN is controlled by the dynamics of the host galaxy. The motivation for introducing $t_{\mathrm{loss}}$ is that the gas accretion rate should also be regulated by the physics that governs the dynamics of gas around the SMBH, as well as the dynamics of the host galaxy. We assume that $t_{\mathrm{loss}}$ depends both on the SMBH mass $M_{\mathrm{BH}}$ and $\Delta M_{\mathrm{BH}}$:
\begin{equation}
t_{\mathrm{loss}} = t_{\mathrm{loss,0}} \left( \frac{M_{\mathrm{BH}}}{{\rm M}_{\odot}} \right)^{\gamma_{\mathrm{BH}}} \left( \frac{\Delta M_{\mathrm{BH}}}{{\rm M}_{\odot}} \right)^{\gamma_{\mathrm{gas}}},
\end{equation}
where $t_{\mathrm{loss,0}}$, $\gamma_{\mathrm{BH}}$, and $\gamma_{\mathrm{gas}}$ are free parameters, and are set to 1~Gyr, 3.5, and $-4.0$, respectively. Based on this model, the dominant term in $t_{\mathrm{acc}}$ is $t_{\mathrm{loss}}$ for most AGN with low-luminosity and/or at low redshifts, as shown in \citet{Shirakata2019} (see their fig.~7). Due to this term, the number density of AGN with low luminosity increases at $z\la 1.5$.

The gas accretion on to a SMBH leads to AGN activity. The Eddington luminosity is defined as $L_{\mathrm{Edd}} \equiv 4\pi G M_{\mathrm{BH}} m_{\mathrm{p}} c / \sigma_{\mathrm{\scalebox{0.5}{T}}}$, where $c$ is the speed of light, $G$ is the gravitational constant, $m_{\mathrm{p}}$ is the proton mass, and $\sigma_{\mathrm{\scalebox{0.5}{T}}}$ is the Thomson scattering cross-section. With this, the Eddington accretion rate is defined as $\dot{M}_{\mathrm{Edd}} \equiv L_{\mathrm{Edd}}/c^2$. Taking into account the effect of `photon trapping' (e.g. \citealt{Ohsuga2005}) when a super-Eddington accretion occurs, we adopt the following relation for the AGN bolometric luminosity, $L_{\mathrm{bol}}$, normalised by the Eddington luminosity, $\lambda_{\mathrm{Edd}} \equiv L_{\mathrm{bol}} / L_{\mathrm{Edd}}$:
\begin{equation}
\lambda_{\mathrm{Edd}} = \left[ \frac{1}{1+3.5 \{ 1+\tanh(\log (\dot{m} / \dot{m}_{\mathrm{crit}})) \} }  + \frac{\dot{m}_{\mathrm{crit}}}{\dot{m}}  \right]^{-1},
\end{equation}
where $\dot{m}$ is the accretion rate normalised by the Eddington rate, $\dot{m} \equiv \dot{M}_{\mathrm{BH}} / \dot{M}_{\mathrm{Edd}}$. This form is based on \citet{Kawaguchi2003}. This kind of suppression of $\lambda_{\mathrm{Edd}}$ at high $\dot{m}$ is described by the slim disc solution (e.g. \citealt{Mineshige2000}). We set $\dot{m}_{\mathrm{crit}}=10.0$ (\citealt{Shirakata2019}). Further, we use the bolometric correction from \citet{Marconi2004} to obtain the hard X-ray (2-10~keV) and $B$-band luminosity of AGN.
We then use
\begin{equation}
M_{\mathrm{UV}} = M_{B} + 0.85,
\end{equation}
to derive the UV magnitude. This equation is obtained by assuming the template SED presented in \citet{Kawaguchi2001}. For comparison, we also use \citet{Duras2020} bolometric correction to obtain the AGN UV luminosity function in Section~\ref{sec:discussion}.

Using the fiducial parameters in \citet{Shirakata2019}, our new model with the Uchuu simulation reproduces a number of key observed galaxy statistics, including the local mass function of neutral hydrogen, the local SMBH mass function, and the high-z $K$-band LF up to $z\sim 3$  (\citealt{Makiya2016}; \citealt{Shirakata2019}), except for the faint end of the $K$-band and $r$-band LFs of local galaxies. The model slightly under-predicts the number densities at magnitudes fainter than the $M_{*}$ of these LFs. We re-calibrate every parameter by hand and find that the weakening the strength of the supernova (SN) feedback makes the fit better without changing any other statistics significantly. For this we take $\alpha_{\mathrm{hot}} = 3.62$ instead of 3.92 in \citet{Shirakata2019}. Our model also reproduces local scaling relations, such as the Tully–Fisher relation, the size–magnitude relation of spiral galaxies, and the relation between bulge and black hole mass. $\nu^2$GC is a mature model, and has already been used in the study of SMBHs and AGN (\citealt{Enoki2014}; \citealt{Oogi2016, Oogi2017}; \citealt{Shirakata2016}).

\subsection{Uchuu-\texorpdfstring{$\nu^2$}{ν\^2}GC galaxy and AGN statistics}
\label{subsec:model_validity}

Firstly, we briefly present our galaxy statistics to set a baseline for the subsequent AGN results. The $\nu^2$GC model now runs on the Uchuu simulation, however previous work instead used the $\nu^2$GC simulation (\citealt{Ishiyama2015}). This difference required us to re-calibrate the model parameters desclibed in Section~\ref{subsec:model}. The resulting model parameter values are identical to \citet{Shirakata2019}, except for $\alpha_{\mathrm{hot}} = 3.62$ instead of 3.92.

\begin{figure}
    \begin{center}
      \includegraphics[width=1.0\columnwidth]{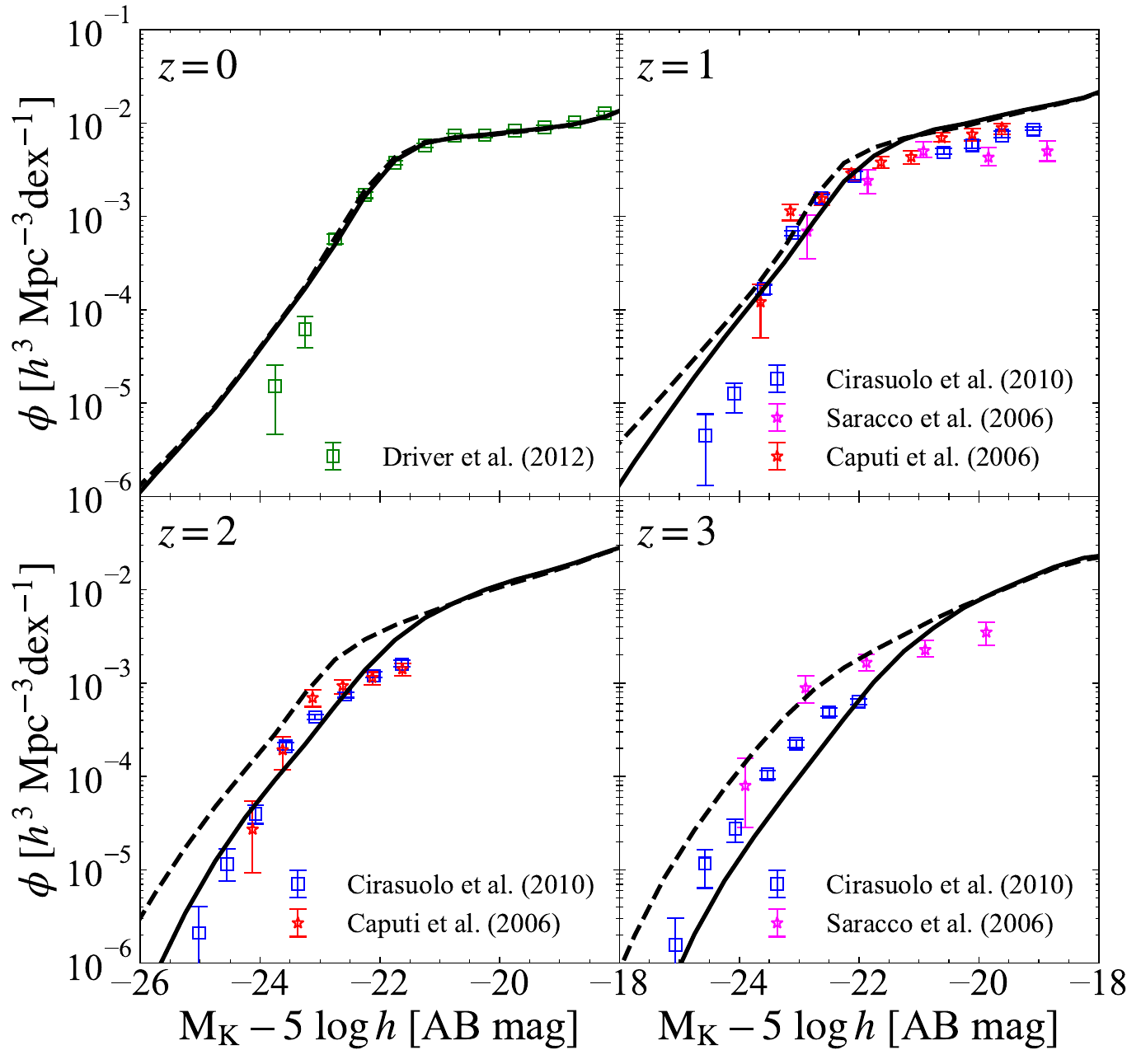}
    \end{center}
  \caption{Rest-frame LFs in the $K$-band at redshift 0, 1, 2, and 3. The solid and dashed lines show the Uchuu-$\nu^2$GC LF with and without dust extinction, respectively. Open green squares at $z=0$ are the observations from \citet{Driver2012}, open red stars at $z=1$ and 2 are from \citet{Caputi2006}, open magenta stars at $z=1$ and 3 are from \citet{Saracco2006}, and open blue squares are from \citet{Cirasuolo2010}.
  }
  \label{fig:kbandLF}
\end{figure}

In Fig.~\ref{fig:kbandLF}, we show the $K$-band galaxy LFs at $z=0,1,2$, and 3. We compare our results with relevant observational data at these redshifts, as labeled. The model $K$-band LF shows good agreement with the observations up to $z=3$, although the effect of dust attenuation becomes too strong at $z=3$. This over-attenuation may be partly due to the estimation of galaxy sizes, which are used to calculate dust surface densities to determine the attenuation in $\nu^2$GC (\citealt{Makiya2016}). A similar tendency to this over-attenuation on the $K$-band has been seen in other semi-analytic models (see fig.~8 in \citealt{Somerville2012} and fig.~7 in \citealt{Lacey2016}). We will address the treatment of dust attenuation in future work. We note that some other studies have also explored the $K$-band LF with semi-analytic models (e.g. \citealt{Henriques2011}; \citealt{Lagos2019}).

\begin{figure}
    \begin{center}
      \includegraphics[width=1.0\columnwidth]{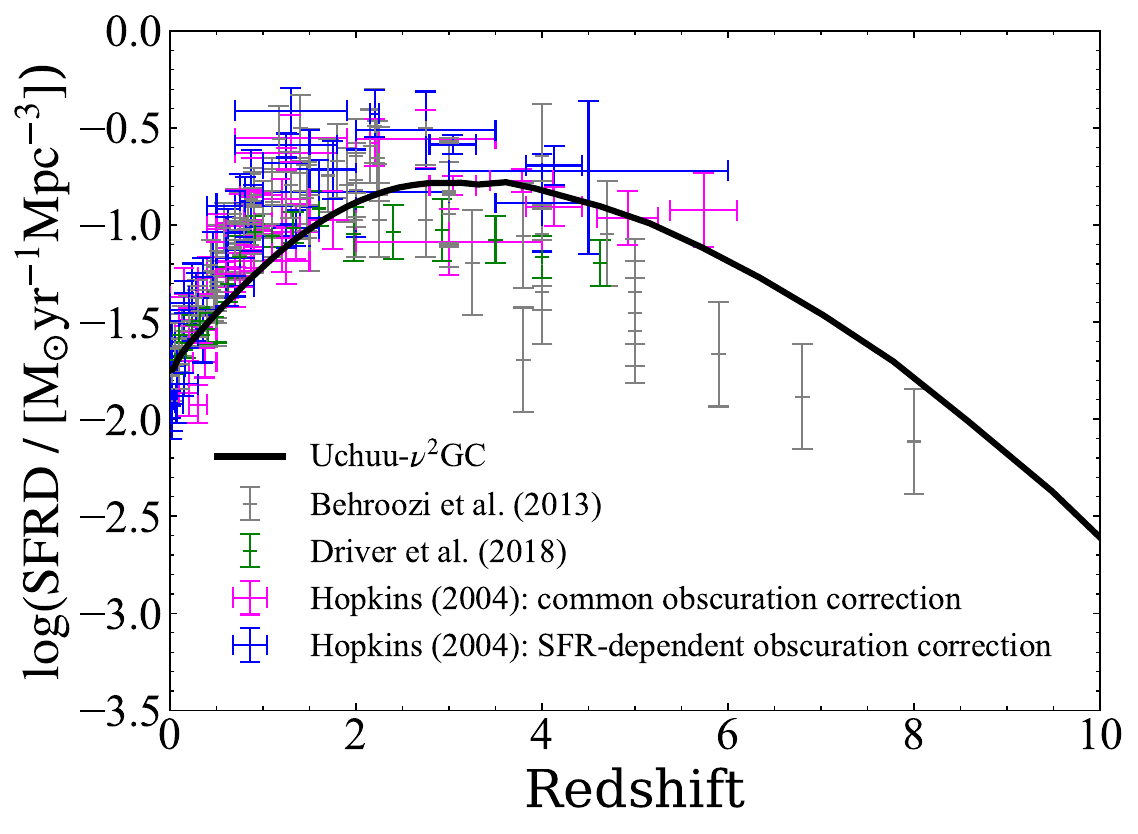}
    \end{center}
  \caption{Cosmic SFR density as a function of redshift. The solid line is the result obtained from Uchuu-$\nu^2$GC. Gray crosses with error bars are a compilation of observational estimates from \citet{Behroozi2013c}. Green data are from \citet{Driver2018}. Magenta (blue) data are from \citet{Hopkins2004} with a common (SFR-dependent) obscuration correction. The model SFRs are converted to a Salpeter initial mass function (IMF) from Chabrier IMF by multiplying by a factor of 1.8.
  }
  \label{fig:csfr}
\end{figure}

Fig.~\ref{fig:csfr} shows the cosmic star formation rate (SFR) density. The overall shape of the redshift evolution is almost consistent with observations, while our model slightly under-predicts the SFR density at $z\sim1-2$. Exploring the reasons for this in the model, we find that disc instabilities strongly shape our SFR density at $z\lesssim3$. With them, central black hole growth is enhanced relative to the model without disc instabilities, and the resultant AGN feedback suppresses the SFR density. Other statistics in the model, such as the black hole mass function in the local Universe and the black hole mass--bulge mass relation, are also in well agreement with recent observational data. Further model galaxy results can be found in \citet{Makiya2016} and \citet{Shirakata2019}.

\begin{figure}
    \begin{center}
      \includegraphics[width=1.0\columnwidth]{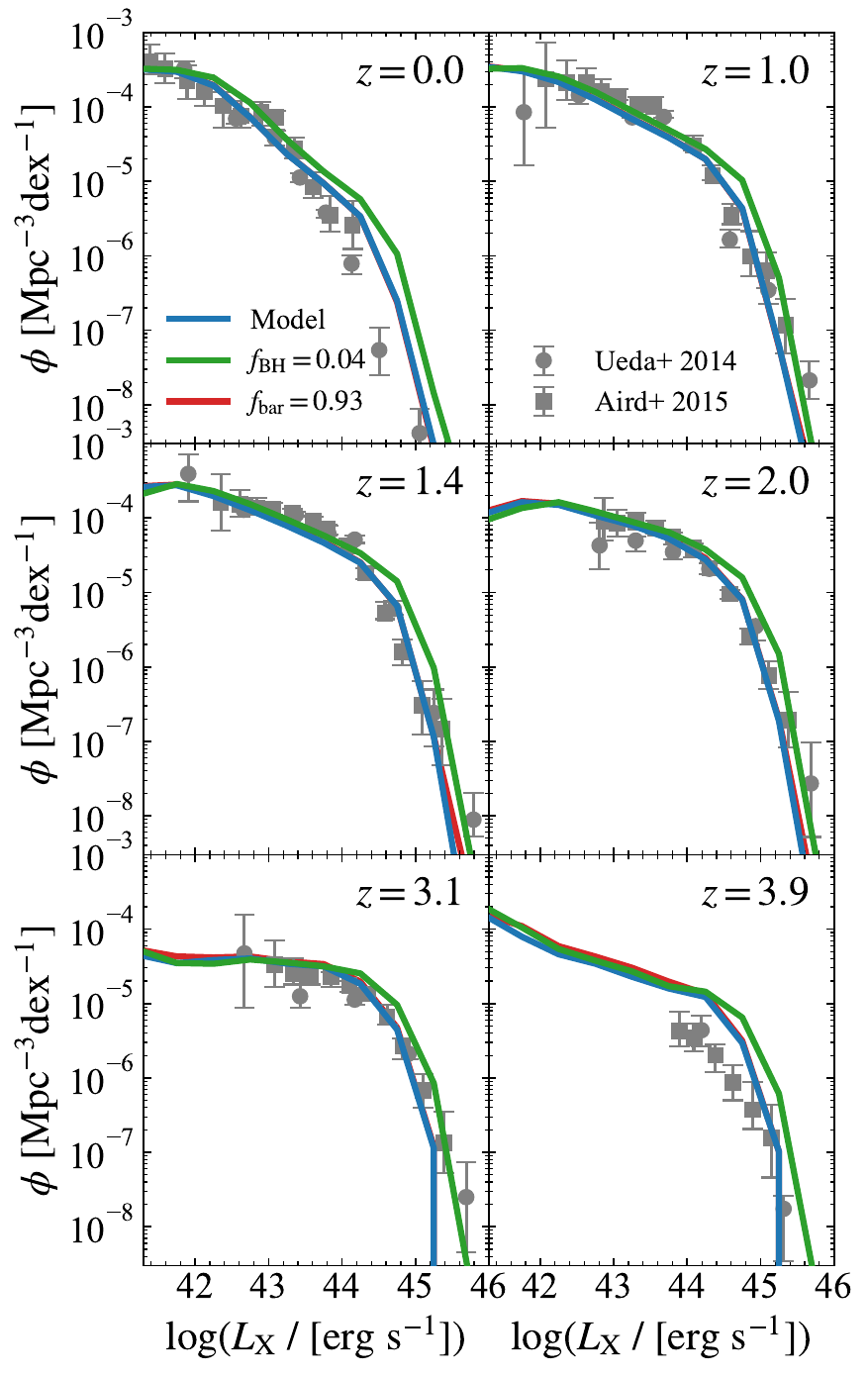}
    \end{center}
  \caption{AGN hard X-ray LFs from $z=0$ to $z=4$. Blue solid lines show the result from Uchuu-$\nu^2$GC. Gray data points with error bars show the observational results from \citet{Ueda2014} and \citet{Aird2015}.
  }
  \label{fig:xlf_ref}
\end{figure}

\begin{figure}
    \begin{center}
      \includegraphics[width=1.0\columnwidth]{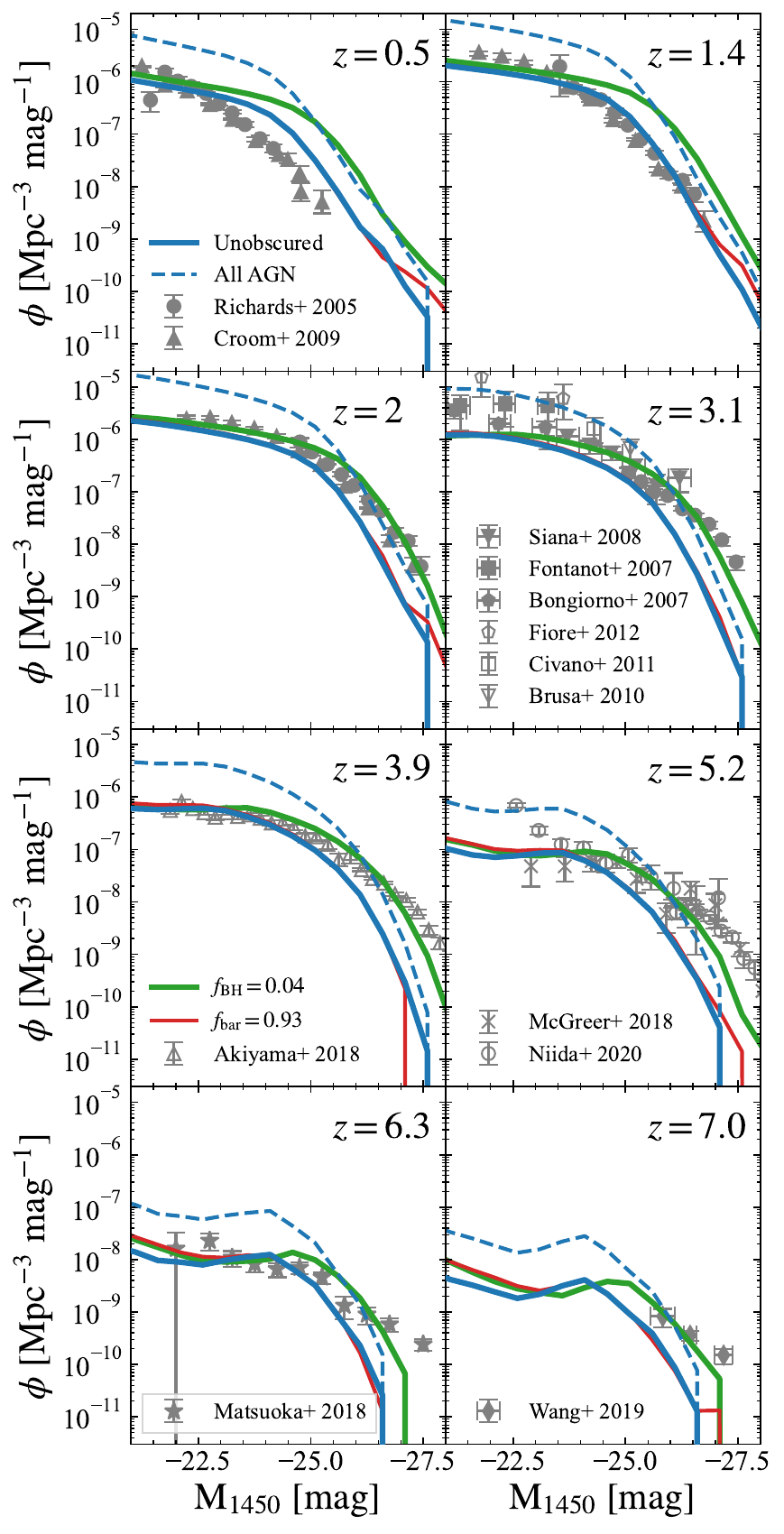}
    \end{center}
  \caption{AGN UV LFs from $z=0$ to $z=7$. Blue solid lines show the result from Uchuu-$\nu^2$GC. Blue dashed lines show the same but without dust obscuration, i.e., the intrinsic AGN LF. Green (red) lines show the result of our model with $f_{\mathrm{BH}}=0.04$ ($f_{\mathrm{bar}}=0.93$). Gray data points with error bars show the observational results from \citet{Richards2005}, \citet{Croom2009}, \citet{Siana2008}, \citet{Fontanot2007}, \citet{Bongiorno2007}, \citet{Fiore2012}, \citet{Civano2011}, \citet{Brusa2010}, \citet{Akiyama2018}, \citet{Niida2020}, \citet{Matsuoka2018}, and \citet{Wang2019}.
  }
  \label{fig:uvlf_ref}
\end{figure}

The next set of figures illustrate how Uchuu-$\nu^2$GC reproduces the observed LF of AGN over a wide redshift range, $0<z<6$ (\citealt{Shirakata2019}). Fig.~\ref{fig:xlf_ref} compares the intrinsic AGN hard {\it X}-ray LFs of Uchuu-$\nu^2$GC and observations from $z=0$ to $z=4$. Fig.~\ref{fig:uvlf_ref} shows the AGN UV LFs from $z=0$ to $z=7$. While our model broadly reproduces the observed AGN UV LF at $\mathrm{M}_{1450} \gtrsim -25$, it under-predicts the bright-end. Figs.~\ref{fig:xlf_ref} and \ref{fig:uvlf_ref} also show the results of two model variants, one with $f_{\mathrm{BH}}=0.04$ and one with $f_{\mathrm{bar}}=0.93$ for comparison. It seems that the model with $f_{\mathrm{BH}}=0.04$ is in more agreement with the observations at $z\gtrsim2$. Since our fiducial model assumes that $f_{\mathrm{BH}}$ is redshift independent, we have set $f_{\mathrm{BH}}=0.02$ to match the LFs at $z<2$ and the observed correlation between the masses of host bulges and their SMBHs at $z=0$ (\citealt{McConnell2013}). We have also verified that the predicted two-point correlation function of AGN is also consistent with current observations (\citealt{Oogi2020}).

\begin{figure}
    \begin{center}
      \includegraphics[width=1.0\columnwidth]{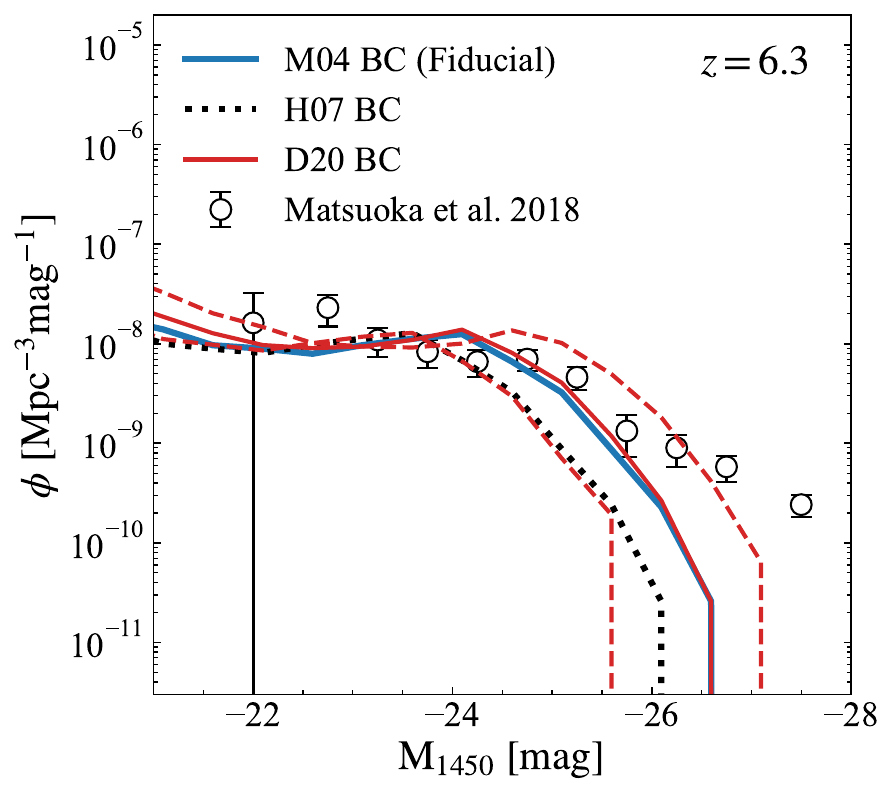}
    \end{center}
  \caption{AGN UV LFs at $z=6.3$ with different optical bolometric corrections, which are the ratios between the bolometric luminosity and the luminosity in the 4400\AA\ band. The blue line shows the result from \citet{Marconi2004}, which we adopt in our fiducial model. The black dotted line shows the result of \citet{Hopkins2007}. Red lines show the the result of \citet{Duras2020}: the the best-fit (solid) and the scatter (dashed).
  }
  \label{fig:lf_bol_corr}
\end{figure}

In this paper we have assumed the empirical SED of AGN which follows a power law in the wavelength range from UV to infrared (\citealt{Marconi2004}) as well as their bolometric correction. Indeed, the UV magnitude of AGN depends on bolometric correction, which can introduce a large uncertainty in the prediction of the LF. We have used two different bolometric corrections from \citet{Hopkins2007} and \citet{Duras2020} to address its effect on the predicted AGN UV LF. This revealed how the bolometric correction affects the bright end of the LF. In Fig.~\ref{fig:lf_bol_corr}, we show the effect of the bolometric correction on the AGN LF at $z=6.3$. The normalization of the bright end varies by $\sim 1$~dex within the scatter found by \citet{Duras2020}. Taking into account the scatter of the bolometric correction, the uncertainties of this correction are larger than those derived for the runs with the different parameters. This leads to different expected numbers of bright AGN for future surveys. Future work will further investigate the impact of bolometric correction on the AGN SED and the resulting detectable number of AGN.

In this work, we focus on AGN with hard {\it X}-ray luminosity $L_{2\mathchar`-10\mathrm{keV}}$ larger than $10^{41}~\mathrm{erg} \ \mathrm{s}^{-1}$, which corresponds to a bolometric luminosity $L_{\mathrm{bol}} \sim 40^{42}~\mathrm{erg} \ \mathrm{s}^{-1}$. The corresponding UV magnitude depends on the dust obscuration. To account for dust obscuration, we use the `observable fraction' introduced by \citet{Shirakata2019}. For intrinsic AGN UV LFs obtained with our model, \citet{Shirakata2019} have empirically determined the observable fraction that fits the observed AGN UV LF (see \citealt{Shirakata2019} and their equation (21) for details).

\subsection{Quantification of cosmic variance}
\label{subsec:model_cv}

Here we describe how we quantify the cosmic variance of the AGN samples in our Uchuu-$\nu^2$GC catalog. The \emph{relative} cosmic variance $\sigma_{\mathrm{cv}}$ can be defined as follows (\citealt{Somerville2004}; \citealt{Bhowmick2020}):
\begin{equation}
\label{eq:cv}
\sigma_{\mathrm{cv}}^2 = \frac{\langle N^2 \rangle - \langle N \rangle^2 - \langle N \rangle}{\langle N \rangle^2},
\end{equation}
where $N$ is the number of AGN in each subvolume and $\langle N \rangle$ is the mean number of AGN across all subvolumes. The first two terms in equation (\ref{eq:cv}) represent the total field-to-field variance $\sigma_{\mathrm{tot}}^2$ which includes the contribution from the cosmic and Poisson variances, following the convention in \citet{Bhowmick2020}. The third term represents the Poisson variance which is subtracted to obtain $\sigma_{\mathrm{cv}}$.

To evaluate $\sigma_{\mathrm{cv}}$, a procedure using the two-point correlation function is usually used (e.g. \citealt{Somerville2004}; \citealt{Trenti2008}; \citealt{Moster2011}). \citet{Bhowmick2020} have primarily used this method to determine $\sigma_{\mathrm{cv}}$ for $z>7$ galaxies. In this paper, we determine $\sigma_{\mathrm{cv}}$ by directly computing $\langle N \rangle$ and $\langle N^2 \rangle$ from the AGN sample in subvolumes, taking advantage of the Uchuu simulation's large volume. Considering the areas of future surveys from telescopes such as {\it JWST} ($\sim 46~\mathrm{arcmin}^2$ and $\sim190~\mathrm{arcmin}^2$), and  Euclid-deep ($\sim 40~\mathrm{deg}^2$), the Uchuu-$\nu^2$GC volume is large enough to extract a number of independent AGN subvolume samples, needed for calculating the cosmic variance.

In this study, we use a single snapshot to evaluate the cosmic variance. choosing the $z$-axis of the simulation box, to represent the redshift width of mock surveys. Table~\ref{tbl:survey2} describes the subvolumes extracted having survey area $A$ and redshift width $\Delta z$. We focus on redshifts $z = 5.72, 6.34, 7.03, 7.78, 8.58$, and $9.48$, which correspond to the Uchuu snapshots. For each snapshot with redshift $z_i$, we consider the comoving length corresponding $\Delta z = 1$ centred on $z_i$, and extract it along with the $z$-axis direction of the simulation box. We then extract the survey areas $A = 0.01, 0.1, 1, 10, 30$, and 80 $\mathrm{deg}^2$ which have a square geometry with side-length corresponding to $\sqrt{A}$ in the $x-y$ plane of the simulation box. We convert degrees to the comoving lengths at each redshift $z_i$. Table~\ref{tbl:survey} shows the comoving lengths of the subvolumes obtained from our simulation box for each survey area and redshift.

In addition to the cuboid geometry described above, we also consider a cubic geometry. This is because previous studies have shown that the survey geometry affects the cosmic variance of the galaxy LF (\citealt{Moster2011}; \citealt{Bhowmick2020}). For the cubic geometry, we extract cubes with side length $L_{\mathrm{eff}} (= V_{\mathrm{eff}}^{1/3})$, where $V_{\mathrm{eff}}$ is a survey volume corresponding to the cuboid geometry. We test the effect of geometry on the total field-to-field variance in Section~\ref{subsec:expected_var}.


\begingroup
\renewcommand{\arraystretch}{1.0}
\begin{table*}
\caption{Summary of the number of Uchuu-$\nu^2$GC subvolumes corresponding to different survey sizes given the redshift width $\Delta z = 1$.
}
\begin{center}
\begin{tabular}{cccccccccc}\hline\hline
\multicolumn{1}{|c|}{Survey area (deg$^2$)} & \multicolumn{7}{c|}{Number of subvolumes}\\
 & redshift & 3.93 & 5.15 & 5.72 & 6.34  & 7.03 & 7.78 \\ \hline
0.01 & & 215296 & 220500 & 249696 & 274428  & 294912 & 311364 \\
\\
0.1 & & 21316 & 21780 & 24576 & 26908  & 28800 & 30276 \\
\\
1.0 & & 2116 & 2205 & 2400 & 2527  & 2888 & 2916 \\
\\
10.0 & & 196 & 180 & 216 & 252  & 288 & 225 \\
\\
30.0 & & 64 & 45 & 54 & 63  & 72 & 81 \\
\\
80.0 & & 16 & 20 & 24 & 28  & 32 & 36 \\

\hline
\label{tbl:survey2}
\end{tabular}
\end{center}
\end{table*}
\endgroup

\begingroup
\renewcommand{\arraystretch}{1.0}
\begin{table*}
\caption{Summary of the comoving lengths (in units of $h^{-1}\mathrm{Mpc}$) corresponding to different Uchuu-$\nu^2$GC survey sizes given the redshift width $\Delta z = 1$.
}
\begin{center}
\begin{tabular}{cccccccccc}\hline\hline
\multicolumn{1}{|c|}{Survey area (deg$^2$)} & \multicolumn{9}{c|}{Comoving length ($h^{-1}\mathrm{Mpc}$)}\\
 & redshift & 3.93 & 5.15 & 5.72 & 6.34  & 7.03 & 7.78 & 8.58 & 9.48 \\ \hline
0.01 & & 8.6 & 9.5 & 9.8 & 10.1 & 10.4 & 10.7 & 11.0 & 11.3 \\
\\
0.1 & & 27.3 & 30.1 & 31.1 & 32.1 & 33.0 & 34.0 & 34.8 & 35.7 \\
\\
1.0 & & 86.2 & 95.0 & 98.3 & 101.4 & 104.5 & 107.4 & 110.1 & 112.7 \\
\\
10.0 & & 272.6 & 300.6 & 310.9 & 320.8 & 330.3 & 339.5 & 348.1 & 356.5 \\
\\
30.0 & & 472.2 & 520.6 & 538.4 & 555.6 & 572.1 & 558.0 & 602.9 & 617.5\\
\\
80.0 & & 711.1 & 850.1 & 879.3 & 907.2 & 934.2 & 960.3 & 984.5 & 1008.0 \\

\hline

Redshift width & \\
$\Delta z = 1$ & & 491.048 & 353.007 & 309.415 & 271.168 & 237.295 & 207.421 & 182.031 & 158.959\\

\hline
\label{tbl:survey}
\end{tabular}
\end{center}
\end{table*}
\endgroup

\section{Cosmic variance of AGN}
\label{sec:result}

\begingroup
\renewcommand{\arraystretch}{1.0}
\begin{table}
\caption{Results of power-law fits to the cosmic variance of AGN hard {\it X}-ray LFs obtained from Equation \ref{eq:cv_fit}.
}
\begin{center}
\begin{tabular}{cccc}\hline\hline
$L_{\mathrm{X}}$ & $z$ & $\alpha$ & $\Sigma \ (\times 10^{-3})$ \\ \hline
40.25 & 3.9 & -0.39 & 33.48 \\
41.25 & 3.9 & -0.40 & 36.22 \\
42.25 & 3.9 & -0.40 & 43.74 \\
43.25 & 3.9 & -0.44 & 44.70 \\
44.25 & 3.9 & -0.41 & 41.34 \\
\\
40.25 & 5.2 & -0.42 & 45.96 \\
41.25 & 5.2 & -0.46 & 42.50 \\
42.25 & 5.2 & -0.44 & 48.18 \\
43.25 & 5.2 & -0.51 & 51.91 \\
44.25 & 5.2 & -0.34 & 98.33 \\
\\
40.25 & 5.7 & -0.40 & 55.18 \\
41.25 & 5.7 & -0.39 & 58.77 \\
42.25 & 5.7 & -0.54 & 40.34 \\
43.25 & 5.7 & -0.35 & 84.59 \\
44.25 & 5.7 & -0.34 & 116.1 \\
\\
40.25 & 6.3 & -0.40 & 62.84 \\
41.25 & 6.3 & -0.44 & 58.41 \\
42.25 & 6.3 & -0.38 & 60.31 \\
43.25 & 6.3 & -0.33 & 111.6 \\
44.25 & 6.3 & -0.23 & 78.93 \\
\\
40.25 & 7.0 & -0.42 & 67.89 \\
41.25 & 7.0 & -0.29 & 91.85 \\
42.25 & 7.0 & -0.20 & 119.2 \\
43.25 & 7.0 & -0.32 & 104.0 \\
44.25 & 7.0 & -0.27 & 327.5 \\
\\
40.25 & 7.8 & -0.24 & 114.7 \\
41.25 & 7.8 & -0.15 & 162.1 \\
42.25 & 7.8 & -0.12 & 157.8 \\
43.25 & 7.8 & - & - \\
44.25 & 7.8 & -1.03 & 4703 \\
\hline

\label{tbl:cv_fit}
\end{tabular}
\end{center}
\end{table}
\endgroup


\begin{figure*}
    \begin{center}
      \includegraphics[width=170mm]{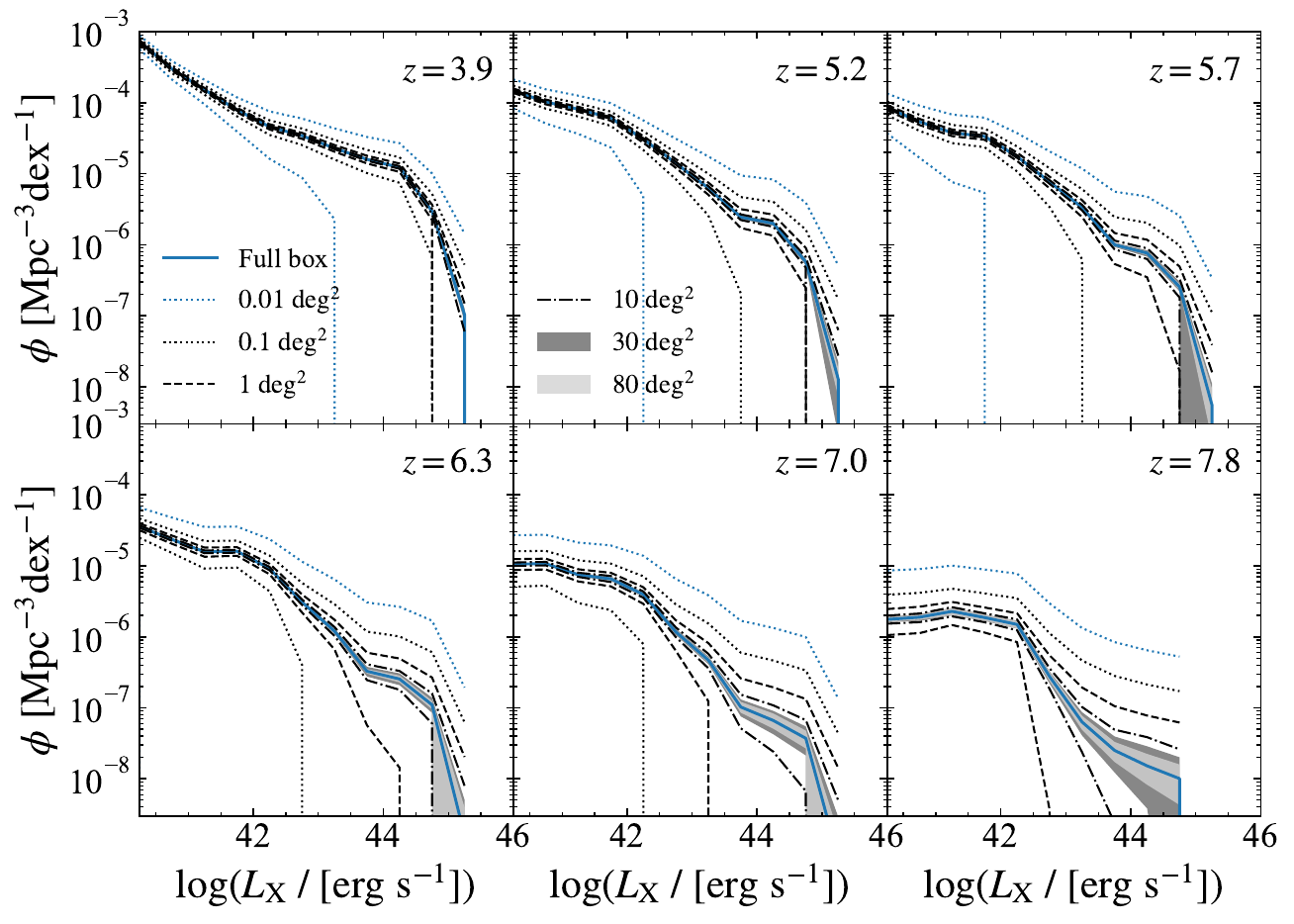}
    \end{center}
  \caption{AGN hard X-ray LFs from Uchuu-$\nu^2$GC for various survey areas (0.01, 0.1, 1, 10, 30, and 80~$\mathrm{deg}^2$) from $z=3.9$ to $z=7.8$. The blue solid line in each panel shows the LF obtained from the full volume of the Uchuu simulation box, $(2~h^{-1}\mathrm{Gpc})^3$. Dashed and dot--dashed lines depict the total field-to-field variance of the LF with 1 and 10~$\mathrm{deg}^2$ survey areas. Light gray and dark gray regions depict those with 30 and 80~$\mathrm{deg}^2$ survey areas, respectively.
  }
  \label{fig:lf_various}
\end{figure*}

\begin{figure*}
    \begin{center}
      \includegraphics[width=170mm]{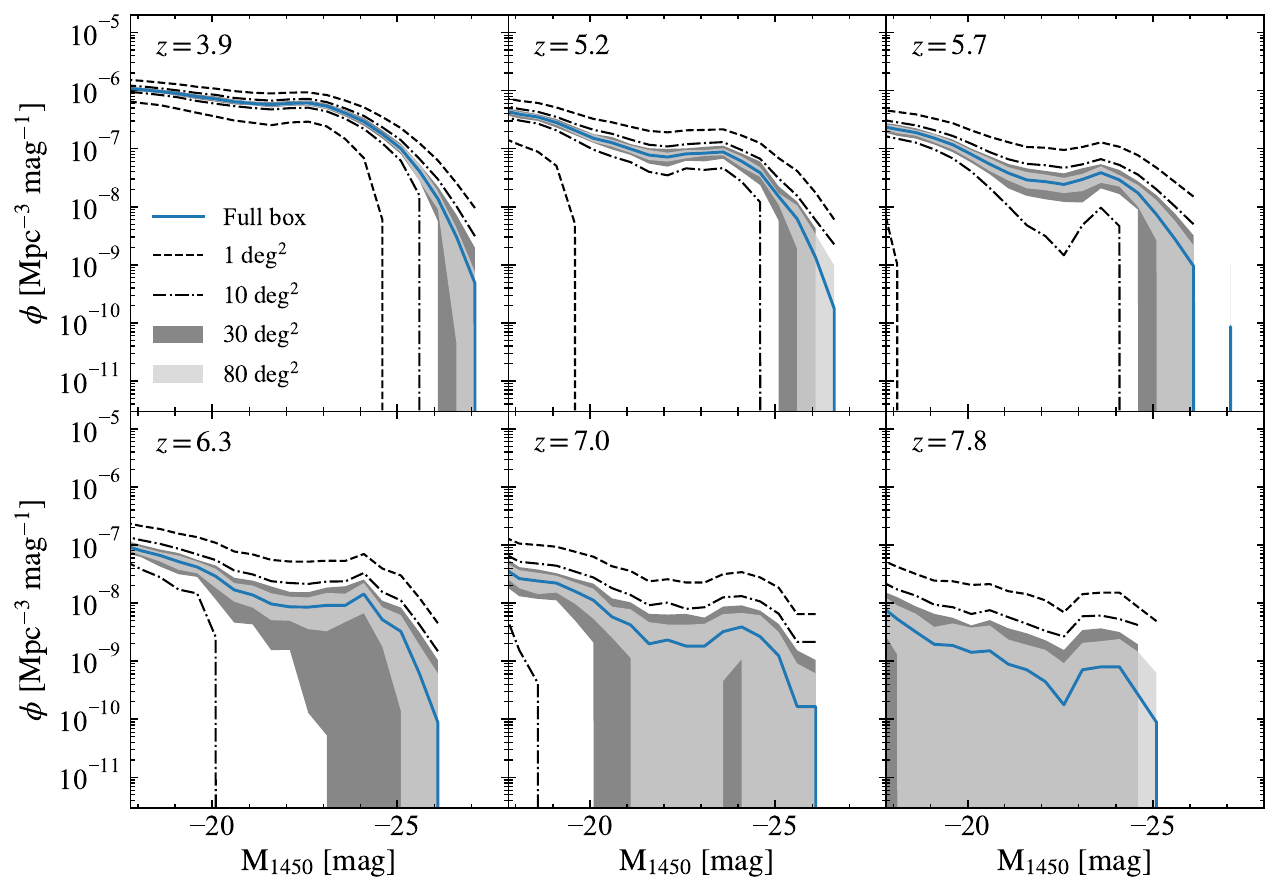}
    \end{center}
  \caption{Same as Fig.~\ref{fig:lf_various}, but for AGN UV LFs.
  }
  \label{fig:uvlf_various}
\end{figure*}

\subsection{AGN LF in each subvolume}
Firstly, in Fig.~\ref{fig:lf_various} we present AGN hard X-ray LFs for different survey areas: $0.01, 0.1, 1, 10, 30$, and 80 $\mathrm{deg}^2$ from $z=3.9$ to 7.8. The shaded regions and bounding lines show $\langle N \rangle (1\pm\sigma_{\mathrm{tot}})$, where $\sigma_{\mathrm{tot}}$ is the total field-to-field variance. In the case of a $1\,\mathrm{deg}^2$ area,  the faint end, $\log(L_{\mathrm{X}} / \mathrm{erg \ s^{-1}}) \lesssim 42$, of the LF has a scatter of $\sim 0.3$ dex and $\sim 0.5$ dex at $z=5.7$ and $z=7.0$, respectively. The bright end, $\log(L_{\mathrm{X}} / \mathrm{erg \ s^{-1}}) \gtrsim 44$, is noisy due to Poisson variance. The spread in the LF decreases with increasing survey area. For 30 and 80 $\mathrm{deg}^2$ surveys, the LFs converge to the LF obtained from the full volume of the Uchuu simulation box, $(2~h^{-1} \mathrm{Gpc})^3$, except at the bright end. As expected given the reduced number of AGN, the scatter is larger at higher redshift in every survey area.

Fig.~\ref{fig:uvlf_various} shows the AGN UV LFs in different survey areas: $1, 10, 30$, and 80 $\mathrm{deg}^2$ from $z=3.9$ to 7.8. While the faint end ($M_{1450} \gtrsim -22$) can be constrained by a 1~$\mathrm{deg}^2$ survey (with $\Delta z \sim 1$) at $z\leq 5.7$, the brighter end cannot, at all redshifts investigated. Using a 10 (30)~$\mathrm{deg}^2$ survey, the LFs for $M_{1450} \gtrsim -24$ can be constrained within 0.3~dex up to $z=6.3$ (7.0). At $z=7.8$, the AGN LF is significantly affected by field variance and cannot be constrained by a 30~$\mathrm{deg}^2$ or smaller survey. At this redshift for the 80~$\mathrm{deg}^2$ survey, the LFs at $M_{1450} \sim -24$ can be constrained within 0.3~dex. Our results show that the AGN UV LF at $M_{1450} \lesssim -26$ is unconstrained by anything smaller than a 80~$\mathrm{deg}^2$ survey.

\begin{figure*}
    \begin{center}
      \includegraphics[width=170mm]{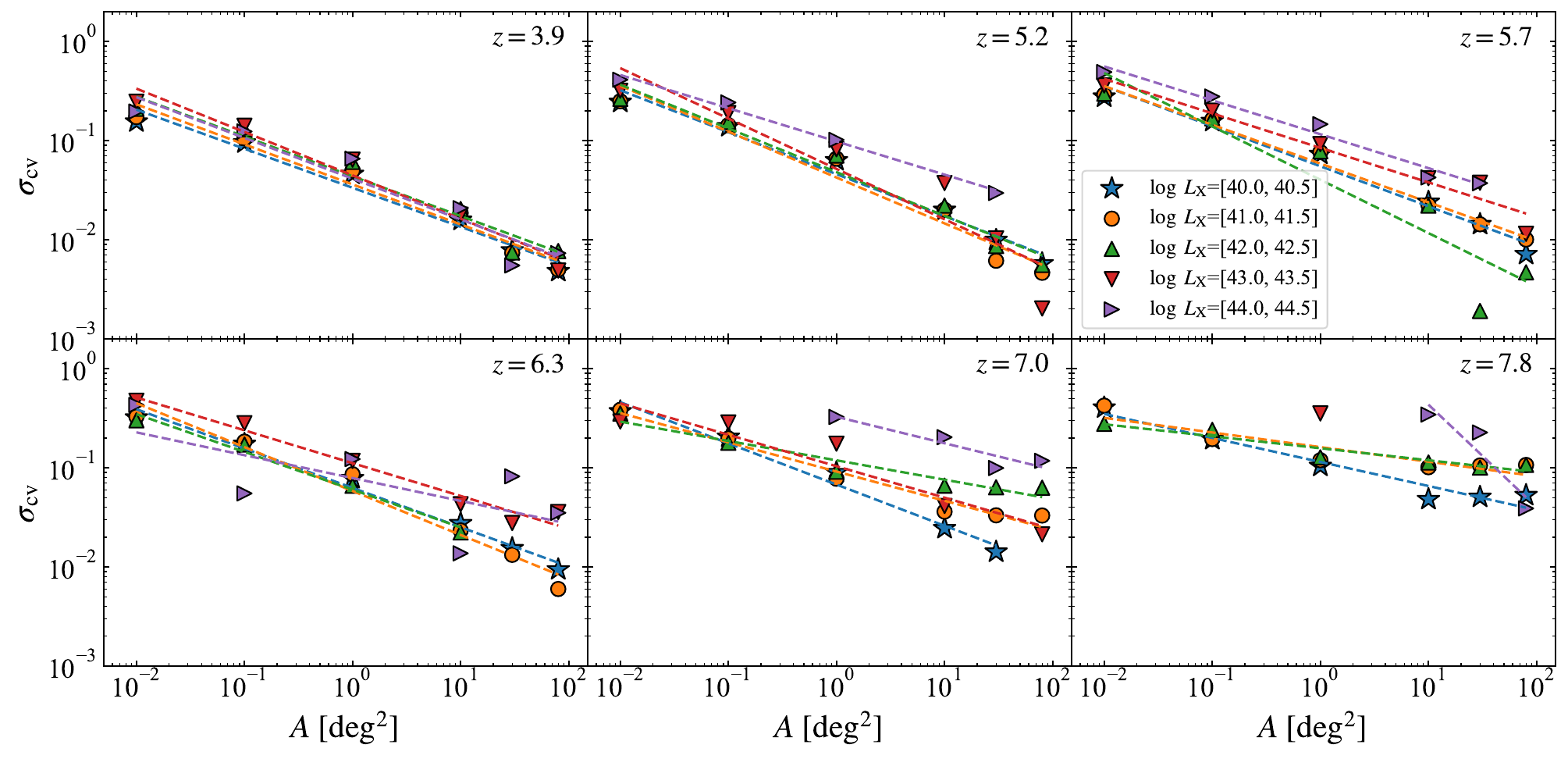}
    \end{center}
  \caption{Cosmic variance as a function of survey area $A$ for AGN samples with four luminosity ranges, $\log (L_{\rm X} / \mathrm{erg~s^{-1}}) =$ [40.0, 40.5] (blue), [41.0, 41.5] (orange), [42.0, 42.5] (green), [43.0, 43.5] (red), and [44.0, 44.5] (purple), at different redshifts. The comoving length of the line-of-sight for each survey area is shown in Table~\ref{tbl:survey}.
  }
  \label{fig:A_sigma}
\end{figure*}

\subsection{Dependence of the cosmic variance on the survey area}

Fig.~\ref{fig:A_sigma} summarises the measured cosmic variance of AGN using $\sigma_{\mathrm{cv}}$ as a function of survey area. Each panel corresponds to a different redshift from $z=3.9$ to $z=7.8$, and shows five luminosity ranges. As seen by the LFs in Fig.~\ref{fig:lf_various}, $\sigma_{\mathrm{cv}}$ decreases with increasing survey area for almost all luminosity ranges. The decreasing trend weakens at high redshift. $\sigma_{\mathrm{cv}}$ ranges from $3 \times 10^{-3}$ to $\gtrsim 0.1$ depending on AGN luminosity and redshift.

The dependence of cosmic variance on survey area is fitted by a power-law function, 
\begin{equation}
\label{eq:cv_fit}
\sigma_{\mathrm{cv}} = \Sigma \ (A / \mathrm{deg}^2)^{\alpha},
\end{equation}
where $\Sigma$ is the normalization and $\alpha$ represents the slope of the power-law. This is the same form as the one presented by \citet{Bhowmick2020}. Table~\ref{tbl:cv_fit} summarizes the best-fitting values of $\Sigma$ and $\alpha$ using the least-squares method for the hard X-ray LFs. Overall, the slope $\alpha$ ranges between $\sim 0.5$ and $\sim 0.1$. These values are roughly consistent with the results of \citet{Bhowmick2020} for $z>7$ galaxies in the \textsc{BlueTides} simulation.

\begin{figure*}
    \begin{center}
      \includegraphics[width=170mm]{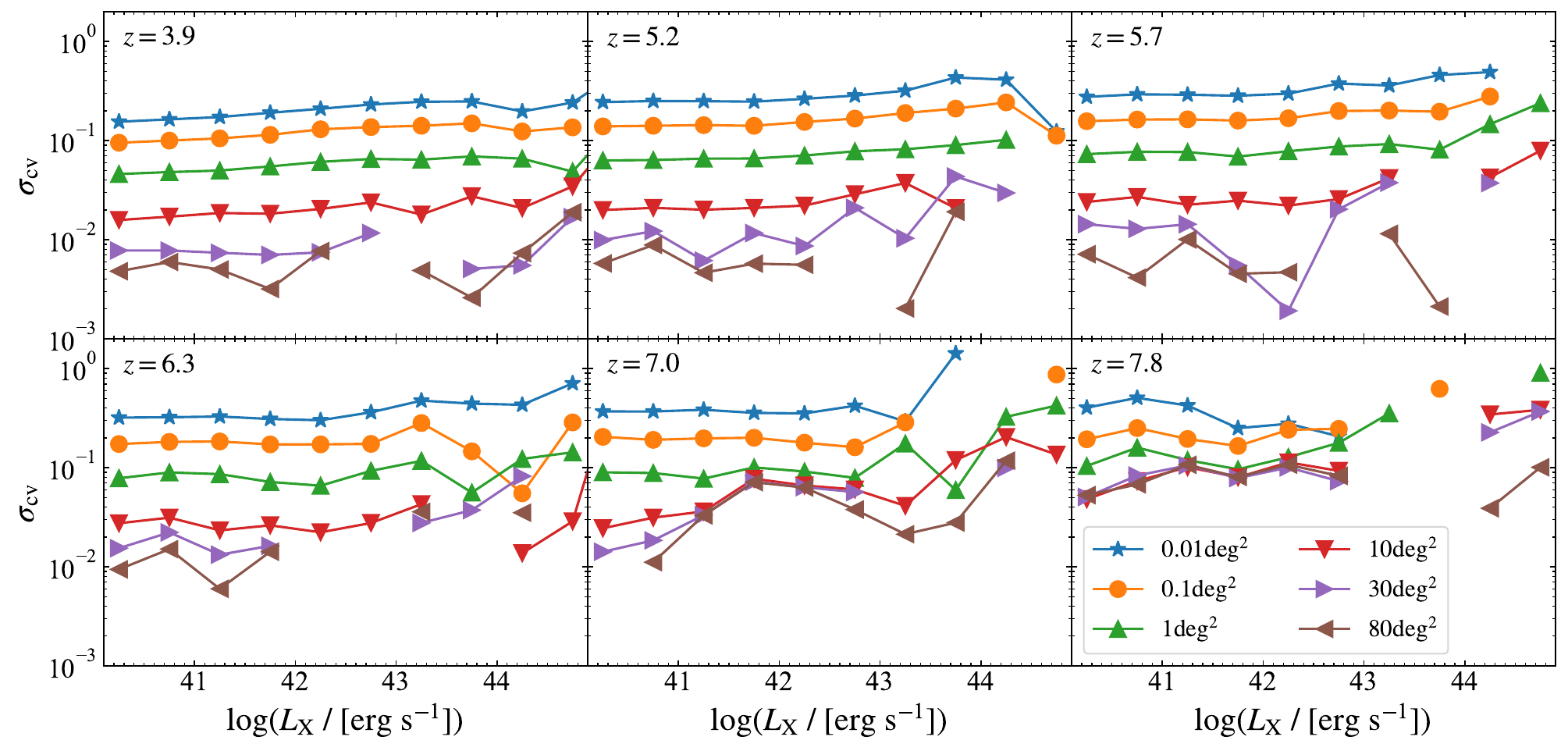}
    \end{center}
  \caption{Cosmic variance as a function of AGN luminosity for various survey areas and at different redshifts. The comoving length of the line-of-sight for each survey area is shown in Table~\ref{tbl:survey}.
  }
  \label{fig:lx_sigma}
\end{figure*}

\subsection{Dependence of the cosmic variance on the AGN luminosity and redshift}

Next, we examine the luminosity dependence of cosmic variance in our model. Fig. \ref{fig:lx_sigma} shows the cosmic variance for different survey areas as a function of AGN luminosity. From $z=3.9$ to 6.3, $\sigma_{\mathrm{cv}}$ only depends weakly on AGN luminosity, in particular for small survey areas (0.01 and 0.1 deg$^2$). This is because the typical DM halo mass in which AGN reside does not monotonically rise with increasing luminosity (\citealt{Oogi2020}). Our result is consistent with the weak luminosity dependence of the observed quasar clustering (e.g. \citealt{Croom2005}; \citealt{Myers2007}; \citealt{Shen2007}; \citealt{Padmanabhan2009}; \citealt{Ross2009}; \citealt{Krolewski2015}). Although the weak luminosity dependence may be due to the limited luminosity range currently probed (e.g. \citealt{White2012}), \citet{He2018} have not found any significant luminosity dependence of the quasar clustering down to the UV magnitude $\mathrm{M_{1450}} \sim -22$ at $z \sim 4$. This lack of the luminosity dependence is in contrast to the results of $\sigma_{\mathrm{cv}}$ for $z>7.5$ galaxies shown by \citet{Bhowmick2020}. For their galaxy sample, brighter galaxies are more strongly clustered.

In Fig. \ref{fig:lx_sigma}, we also see the redshift dependence of $\sigma_{\mathrm{cv}}$. For low-luminosity AGN with $\log (L_{\rm X} / \mathrm{erg~s^{-1}}) \simeq 41$, $\sigma_{\mathrm{cv}}$ gradually increases from $z=3.9$ to $z=7.8$ by a factor of 3-20, depending on the survey area. This trend is partly due to the change in the comoving survey volume. If we assume a fixed survey area and redshift depth, the corresponding comoving survey volume decreases with redshift. This is because the fixed redshift interval is chosen ($\Delta z=1$ in this paper). For example, for a 80 $\mathrm{deg}^2$ survey area and $\Delta z=1$, the comoving survey volume is 0.939~$\mathrm{Gpc}^3$ at $z=3.9$, while it is 0.615~$\mathrm{Gpc}^3$ at $z=7.8$, as can be seen from Table~\ref{tbl:survey}. Another possible reason for the redshift dependence is the change in AGN clustering, which leads to the change in $\sigma_{\mathrm{cv}}$. For high-luminosity AGN with $\log (L_{\rm X} / \mathrm{erg~s^{-1}}) \simeq 44$, $\sigma_{\mathrm{cv}}$ also increases from $z=3.9$ to $z=7.0$ by a factor of 5-15, although the trend is noisy due to the finite number density.

\begin{figure*}
    \begin{center}
      \includegraphics[width=170mm]{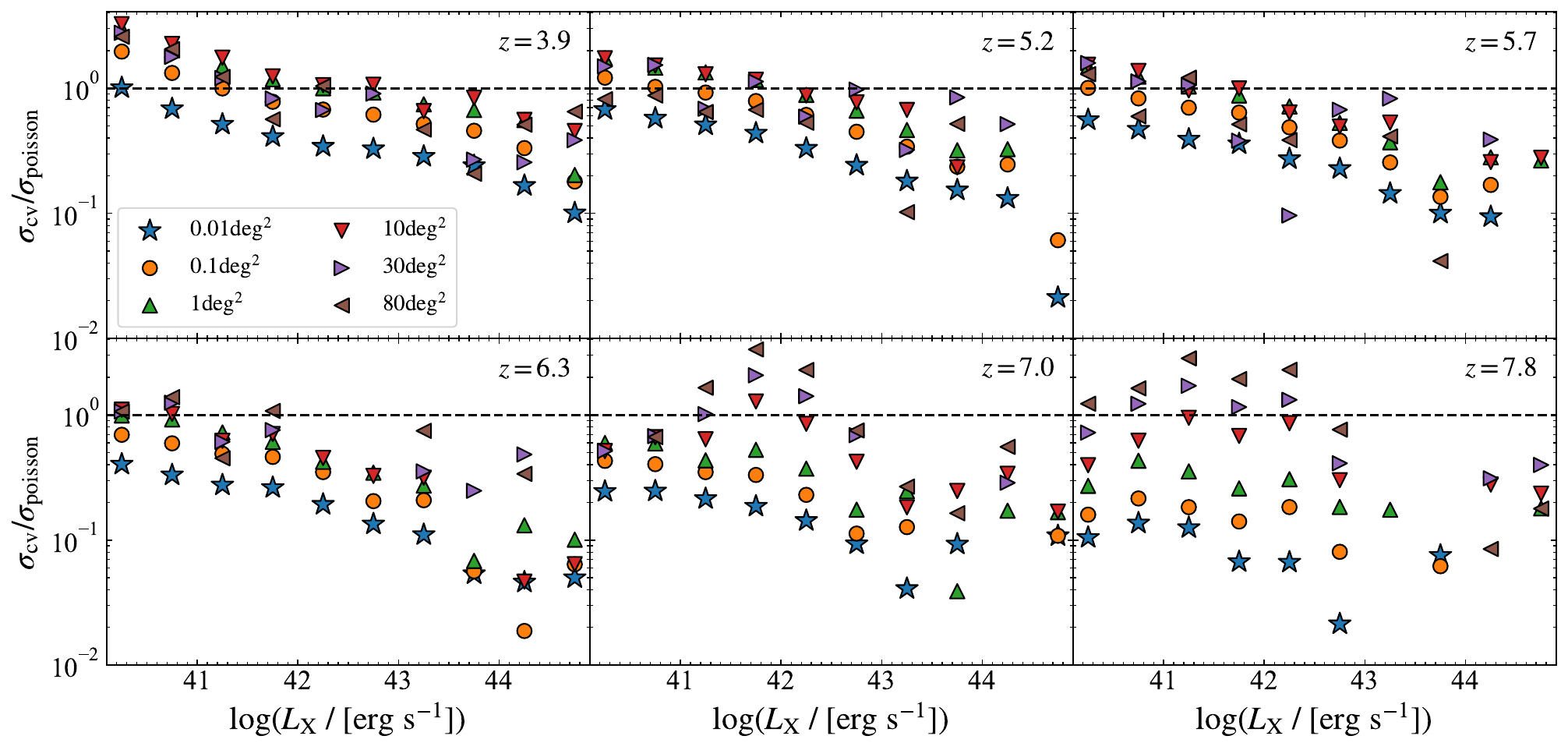}
    \end{center}
  \caption{The ratio of the cosmic variance to Poisson variance as a function of AGN luminosity for various survey areas and at different redshifts. The comoving length of the line-of-sight for each survey area is shown in Table~\ref{tbl:survey}.
  }
  \label{fig:lx_sigma_ratio}
\end{figure*}

Fig.~\ref{fig:lx_sigma_ratio} shows the ratio of the cosmic variance to Poisson variance. In the case of 1 $\mathrm{deg}^2$ and smaller survey areas, the Poisson variance dominates the total variance at all redshifts. In the case of $30$ and 80 $\mathrm{deg}^2$, the cosmic variance dominates for lower-luminosity AGN with $\log(L_{\mathrm{X}} / \mathrm{erg \ s^{-1}}) \lesssim 42$, while the Poisson variance increases for higher-luminosity AGN. This trend weakens at $z\gtrsim 7$. This is because our model predicts the LF with a flatter faint end slope at these redshifts compared to the lower redshifts (see Fig. \ref{fig:lf_various}). In other words, the AGN number density does not depend significantly on the luminosity in $L_{\mathrm{X}} \lesssim 10^{42} \mathrm{erg \ s^{-1}} $ at $z\gtrsim 7$. In this luminosity range, the Poisson variance is comparable to the cosmic variance, and there is no clear trend with luminosity.

It is important to remember that our estimates of the cosmic variance can be affected by uncertainties in the galaxy and AGN formation physics. Here we investigate the effects of SMBH and AGN formation parameters on the total field-to-field variance. We focus on two model parameters, $f_{\mathrm{BH}}$ and $f_{\mathrm{bar}}$. These parameters change the amount of mass growth for each SMBH (see Section 2 and \citealt{Shirakata2019}). While the degree of the variance at the bright end of the LF is slightly different between the two models, the faint ends have similar variance. We will compare the results of the three variants of the model in Section \ref{subsec:future_survey}.

\subsection{Expected variance in current and future surveys}
\label{subsec:expected_var}

\begin{figure*}
    \begin{center}
      \includegraphics[width=170mm]{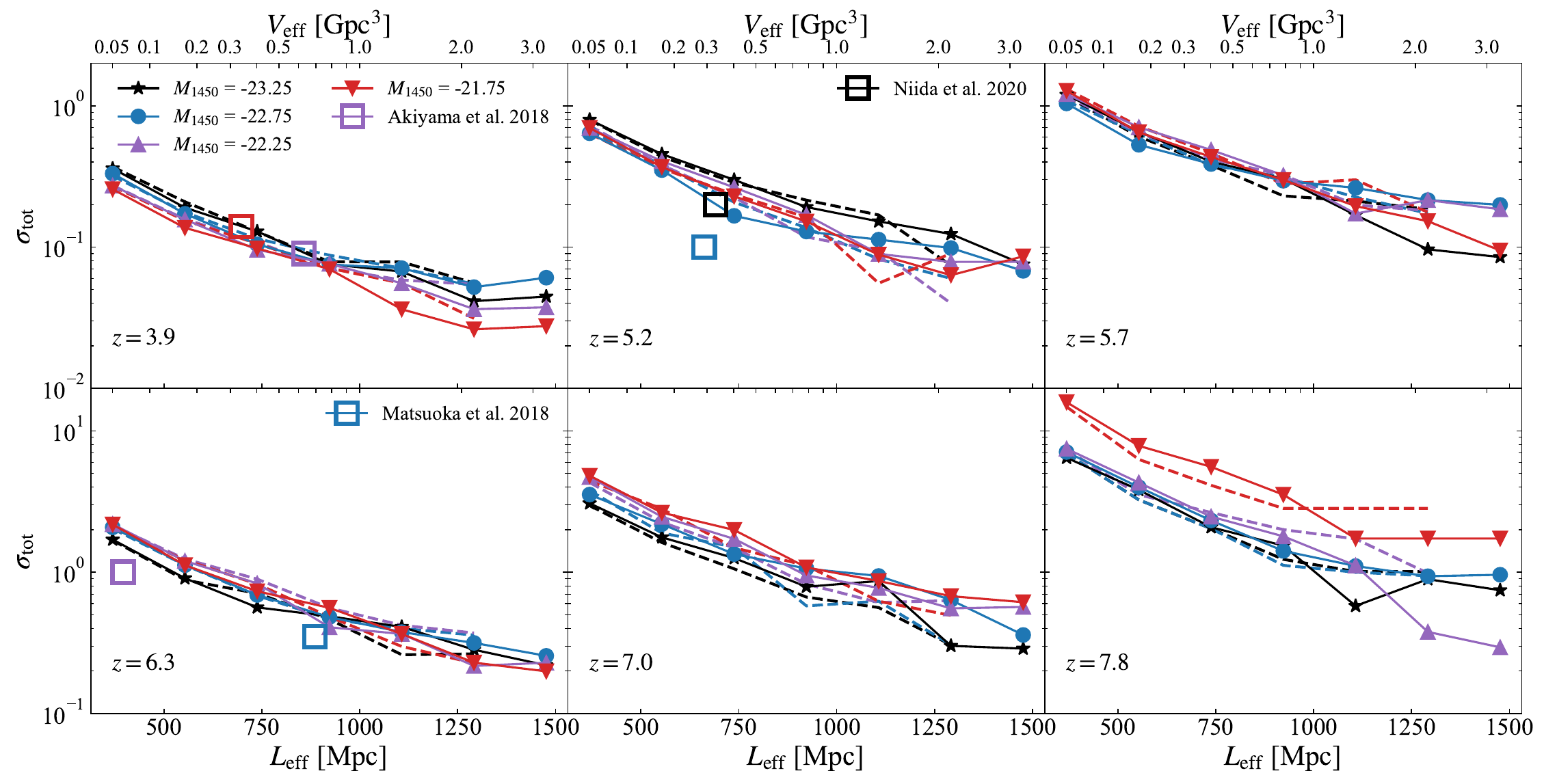}
    \end{center}
  \caption{The total field-to-field variance as a function of effective box size $L_{\mathrm{eff}}$, corresponding to an effective survey volume $V_{\mathrm{eff}}$, for AGN samples in four luminosity ranges, $M_{1450}=$ [-23.5, -23.0] (black), [-23.0, -22.5] (blue), [-22.5, -22.0] (purple), and [-22.0, -21.5] (red), at different redshifts. Solid lines depict the results of cubic geometries, while dashed lines depict those of cuboid geometries with depth $\Delta z = 1$. The two types of geometry have the same survey volume. The horizontal axis on the top depicts the corresponding effective survey volume. We note that the y-axes of the top and bottom panels differ in range. Open squares show the observational results from \citet{Akiyama2018}, \citet{Niida2020}, and \citet{Matsuoka2018}. The colours indicate the range of magnitudes corresponding to the model results.
  }
  \label{fig:Leff_sigma}
\end{figure*}

Our extremely large Uchuu-$\nu^2$GC volume enables us to evaluate the expected field variance for existing and future luminosity function surveys. In this subsection, we focus on the faint end of the AGN UV LF ($M_{\mathrm{UV}} \gtrsim -23$), because this is of importance for studies on the contribution of AGN to reionisation (\citealt{Giallongo2019}; \citealt{Grazian2020}). Fig.~\ref{fig:Leff_sigma} shows the total field-to-field variance $\sigma_{\mathrm{tot}}$ (the first two terms in equation (\ref{eq:cv})), of the faint end of AGN UV LF for several luminosity ranges as a function of effective box size $L_{\mathrm{eff}}$, corresponding to different effective survey volumes $V_{\mathrm{eff}}$. We again consider two types of geometries: cubic and cuboid (see Section \ref{subsec:model_cv}); solid lines correspond to the cubic geometry, while dashed lines correspond to the cuboid geometry.

As expected, $\sigma_{\mathrm{tot}}$ decreases with $V_{\mathrm{eff}}$, like the survey area dependence in Fig.~\ref{fig:A_sigma}. The variances $\sigma_{\mathrm{tot}}$ of different magnitudes are not significantly different in this magnitude range. $\sigma_{\mathrm{tot}}$ increases with redshift. Our results also show that $\sigma_{\mathrm{tot}}$ only weakly depends on survey geometry. This may be because the Poisson variance dominates $\sigma_{\mathrm{tot}}$ in the case of AGN, in contrast to what occurs for galaxies.

We also compare our predictions with current observations in Fig.~\ref{fig:Leff_sigma}. Firstly, from \citet{Akiyama2018} the $V_{\mathrm{eff}}$ of their AGN LF at $z \simeq 4$ is 0.34 (0.63)~$\mathrm{Gpc}^3$ for the magnitude bin of $M_{1450} = -21.875 \ (-22.125)$. At these luminosity ranges, they have estimated the uncertainty of the LF to be $\sim 0.14$ (0.09). For the same luminosity and redshift range, our estimate of the LF scatter, $\sigma_{\mathrm{tot}} \simeq 0.11$ (0.08), is consistent with their values. Secondly, for \citet{Niida2020} the $V_{\mathrm{eff}}$ at $z \simeq 5$ is 0.29 (0.33)~$\mathrm{Gpc}^3$ for the magnitude bin of $M_{1450} = -22.57 \ (-23.07)$, and their corresponding uncertainty is $\sim 0.1$ ($\sim0.2$). In our analysis, $\sigma_{\mathrm{tot}} \simeq 0.24$ (0.35) at the same luminosity and redshift range. Our result is consistent within a factor of two. Finally, for \citet{Matsuoka2018} the $V_{\mathrm{eff}}$ at $z \simeq 6$ is 0.062 (0.694)~$\mathrm{Gpc}^3$ for the magnitude bin of $M_{1450} = -22.0 \ (-22.75)$, and their corresponding uncertainty is $\sim 1$ ($\sim0.35$). In our analysis, $\sigma_{\mathrm{tot}} \simeq 2.0$ ($\simeq 0.5$), again consistent to their observational estimate within a factor of two.

Furthermore, Fig.~\ref{fig:Leff_sigma} shows our predicted uncertainty for LFs that can be obtained from future surveys. For example, at $z = 7.8$ (7.0), $V_{\mathrm{eff}} \gtrsim 2 \ (0.8)~\mathrm{Gpc}^3$ is required to suppress the uncertainty to $\sigma_{\mathrm{tot}} \lesssim 1$. This relation between $L_{\mathrm{eff}}$ and $\sigma_{\mathrm{tot}}$ is needed if one would like to know the required survey area to constrain the AGN LF at a desired level.

\begin{figure*}
    \begin{center}
      \includegraphics[width=170mm]{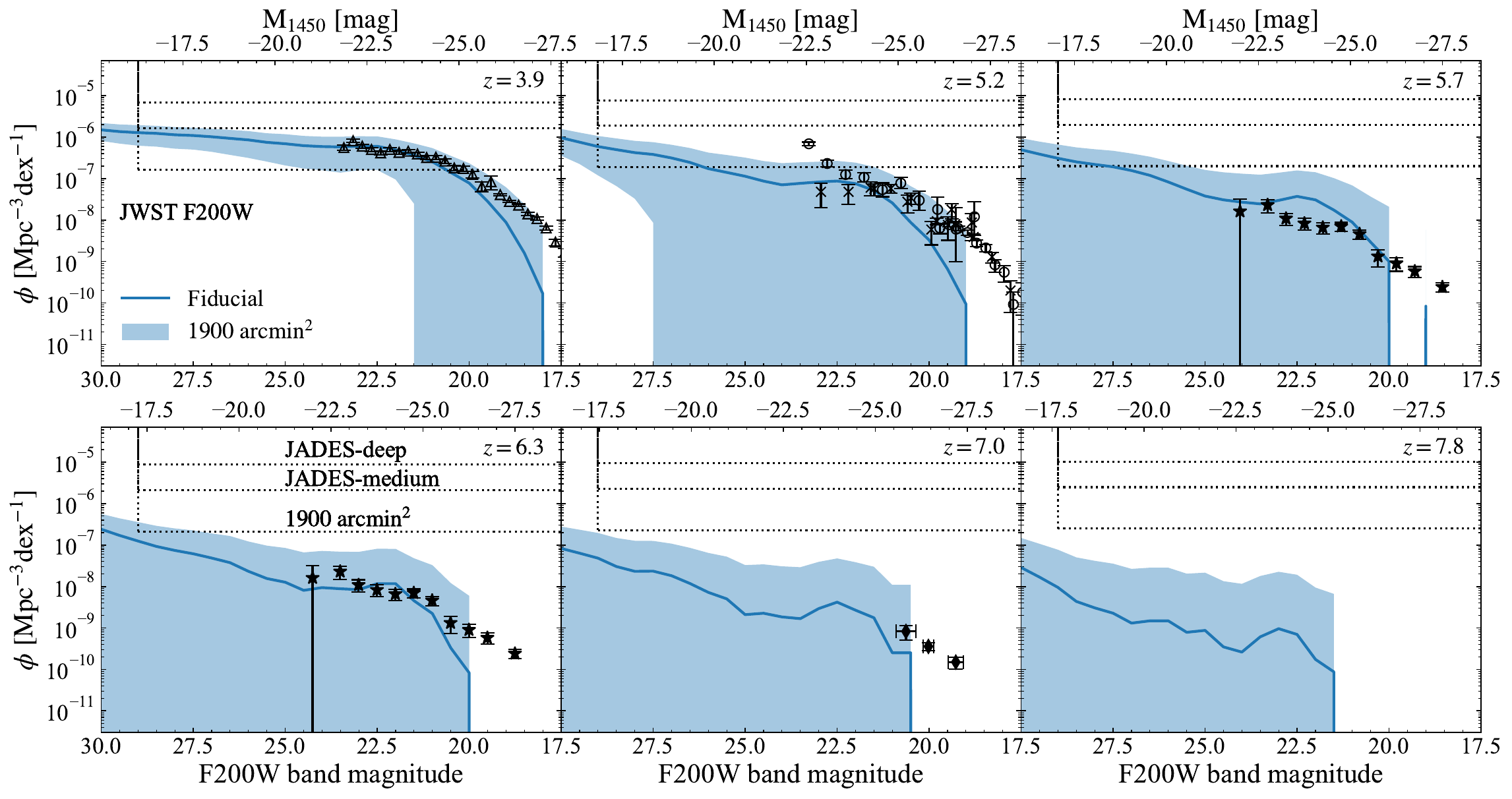}
    \end{center}
  \caption{Predicted AGN LFs using the JWST NIRCam F200W band from $z=3.9$ to $z=7.8$. Blue solid lines show the result for our Uchuu-$\nu^2$GC model with obscuration. Blue shaded regions show the total field-to-field variance of the LF when assuming a $1900~\mathrm{arcmin}^2$ survey area, ten times as large as that of the JADES-medium survey.
  Black data points with error bars are the same observational results as those shown in Fig.~\ref{fig:uvlf_ref}, but converted to apparent magnitude. The horizontal dotted lines depict the number density limit derived from the survey areas of JADES-deep ($46~\mathrm{arcmin}^2$), JADES-medium ($190~\mathrm{arcmin}^2$), and ten times JADES-medium ($1900~\mathrm{arcmin}^2$), from top to bottom. The vertical dotted lines depict the magnitude limit derived from the flux limit adopted in this work ($9.1 \mathrm{nJy}$). Objects above the horizontal line and to the right of the vertical line are detectable. The top horizontal axis shows the rest-frame UV absolute magnitude corresponding to the lower axis apparent magnitude.
  }
  \label{fig:LF_F200W}
\end{figure*}

\begin{figure*}
    \begin{center}
      \includegraphics[width=170mm]{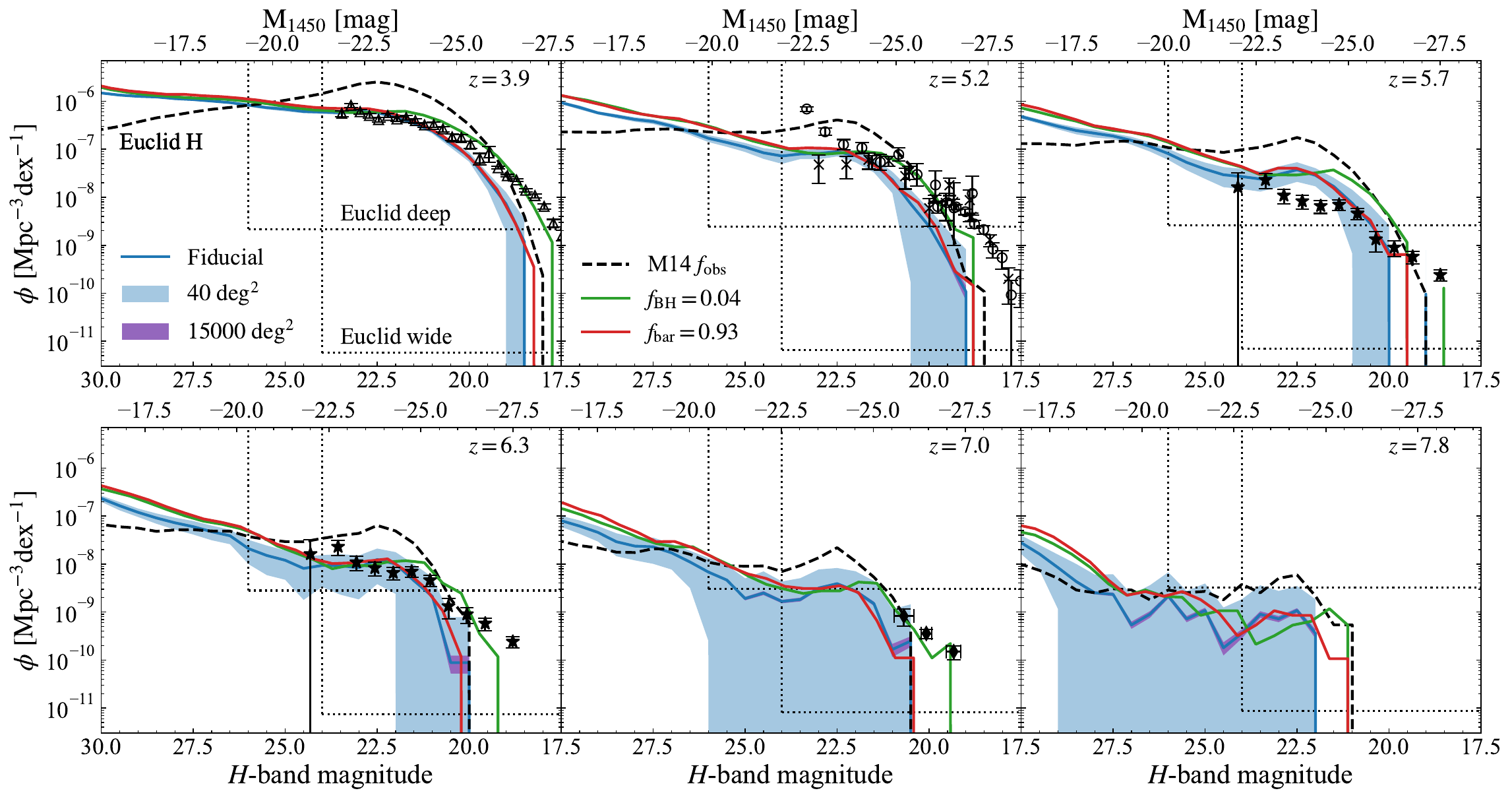}
    \end{center}
  \caption{Same as for Fig.~\ref{fig:LF_F200W}, but for both Euclid H-band surveys. Dark and light gray shaded regions show the total field-to-field variance of the LF when assuming the Euclid deep ($40~\mathrm{deg}^2$) and wide ($15000~\mathrm{deg}^2$) survey area, although the latter region is almost invisible due to the small variance. The variance for $15000~\mathrm{deg}^2$ is estimated by extrapolating from those up to $80~\mathrm{deg}^2$. For comparison, our model variants with the models $f_{\mathrm{BH}}=0.04$ and $f_{\mathrm{bar}}=0.93$ are also plotted (see also Fig.~\ref{fig:uvlf_ref}). Since the field variances of the two models are similar to that of the fiducial model, we omit them. In addition, black dashed lines show the result of the model with the observable fraction of \citet{Merloni2014}. Dotted lines correspond to the magnitude limit and survey area of Euclid deep ($145 \mathrm{nJy}$, $40~\mathrm{deg}^2$) and Euclid wide ($912 \mathrm{nJy}$, $15000~\mathrm{deg}^2$).
  }
  \label{fig:LF_Euclid}
\end{figure*}

\begin{figure*}
    \begin{center}
      \includegraphics[width=170mm]{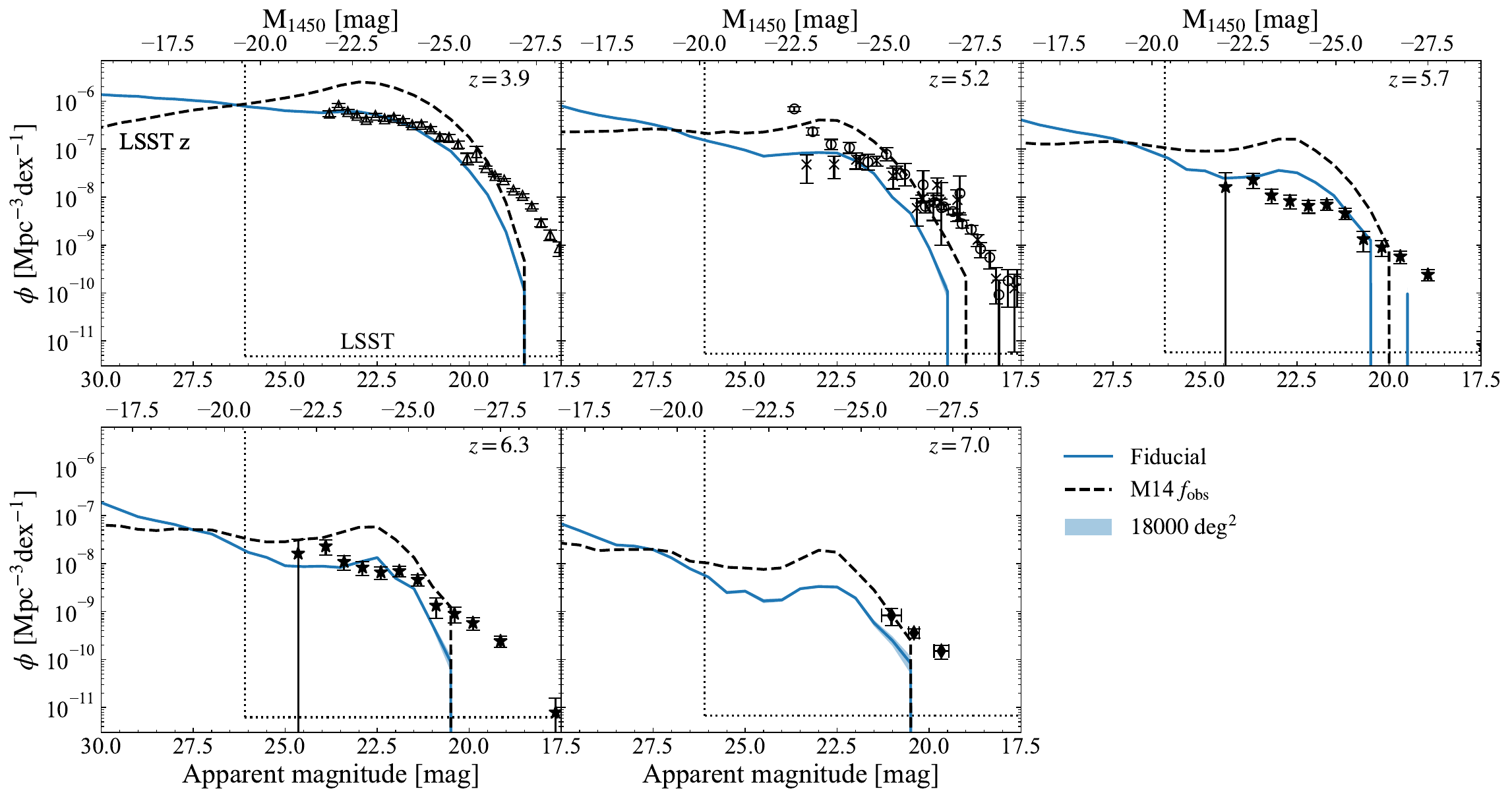}
    \end{center}
  \caption{Same as for Fig.~\ref{fig:LF_F200W}, but in the LSST {\it z}-band. Blue shaded regions show the total field-to-field variance of the LF when assuming the LSST survey area ($18000~\mathrm{deg}^2$), although the region is almost invisible due to the small variance. The variance for $18000~\mathrm{deg}^2$ is estimated by extrapolating from those up to $40~\mathrm{deg}^2$. The dotted lines correspond to the magnitude limit and survey area of the LSST survey ($26.1~\mathrm{mag}$ in the {\it z}-band, $18000~\mathrm{deg}^2$).
  }
  \label{fig:LF_LSST}
\end{figure*}

\subsection{Predictions of AGN luminosity functions for upcoming surveys}
\label{subsec:future_survey}

Telescopes with new instruments, like those onboard the {\it JWST} and {\it Euclid} spacecrafts, and the LSST camera on the Vera Rubin telescope, can expand our understanding of the statistics of high-redshift AGN. To help guide our exploration of the AGN using these new facilities, we present predictions of the AGN LF and expected numbers that will be discovered by future surveys.

{\it JWST}, launched on December 2021, carries onboard the NIRCam instrument observing across the wavelength range 0.7--5 $\mu \mathrm{m}$. With the NIRCam F200W filter it will measure AGN radiation at $z\geq 7$ in the rest-frame ultraviolet and optical wavelengths. Importantly, this radiation is expected to be less affected by the absorption of the inter-galactic medium. To derive the magnitude in the {\it JWST} bands, we convolve the spectral energy distribution (SED) for each AGN with the filter response functions in a set of NIRCam filters\footnote{\url{https://jwst-docs.stsci.edu/jwst-near-infrared-camera/nircam-instrumentation/nircam-filters}}. For an AGN SED, we assume the empirical SED which follows a power law in the wavelength range from UV to infrared (\citealt{Marconi2004}) as well as their bolometric correction. In Fig.~\ref{fig:LF_F200W} we show the predicted AGN LFs in the F200W band observer frame. We assume a flux limit of $9.1~\mathrm{nJy}$, which is the same as that used by \citet{Griffin2020}. For sky coverage, we mimic the JWST Advanced Deep Extragalactic Survey (JADES, \citealt{Williams2018}). We consider three survey areas: 46, 190, and $1900~\mathrm{arcmin}^2$, which correspond to the surveys JADES-deep, JADES-medium, and ten times JADES-medium. We show the mean LF (blue solid lines) and the total field-to-field variance when assuming a $1900~\mathrm{arcmin}^2$ survey area (blue shaded regions). Invisible lower limits of the blue shaded regions in most magnitude ranges mean that no AGN can be detected given the extent of the survey area. Fig.~\ref{fig:LF_F200W} shows that even with $1900~\mathrm{arcmin}^2$ AGN cannot be detected unless the survey area has much more AGN compared to the average.

{\it Euclid}, which is scheduled for launch in 2023, will also have optical and near-IR filters at wavelengths of 0.5--2 $\mu \mathrm{m}$. We present predictions of the AGN LF in the Euclid {\it H}-band in Fig.~\ref{fig:LF_Euclid}. To derive the $H$-band magnitude, we use a set of response functions of {\it Euclid} filters (also including LSST filters) provided by the Euclid collaboration. Two types of surveys are planed: Euclid-deep ($40~\mathrm{deg}^2$ area) and Euclid-wide ($15000~\mathrm{deg}^2$ area). Here, we use the same flux limits adopted by \citet{Griffin2020}, $145~\mathrm{nJy}$ for Euclid-deep and $912~\mathrm{nJy}$ for Euclid-wide. The flux limits and survey areas are shown in Fig.~\ref{fig:LF_Euclid}. In the figure, we show the predicted variance of the AGN LF corresponding to each survey area. For Euclid-deep, we directly calculate the variance extracting regions from our Uchuu-$\nu^2$GC volume. For Euclid-wide, we extrapolate the variance from the results up to $80~\mathrm{deg}^2$. This figure clearly shows that while Euclid-deep constrains the faint end ($\mathrm{M_{1450}}\gtrsim-23$) of the LF, Euclid-wide can also determine the bright end ($\mathrm{M_{1450}}\lesssim-23$).

Furthermore, we study the possibility to distinguish different AGN formation models using the Euclid surveys. In Fig.~\ref{fig:LF_Euclid}, we also show the results of two model variants, one with $f_{\mathrm{BH}} = 0.04$ and one with $f_{\mathrm{bar}} = 0.93$; these are the same models presented in Fig.~\ref{fig:uvlf_ref}. While the amount of variance at the bright end can be different between the two models, the faint end variances are similar. Therefore, the parameters related to SMBH growth and AGN activity in our model do not significantly affect the variance of the AGN LF at the faint end. However, we do find a slightly larger normalisation ($\sim0.3$~dex larger) between the two model variants at the faint end than those for the fiducial model. The model with $f_{\mathrm{BH}}=0.04$ partly reproduces the bright end of the AGN LF at $z \geq 4$ (see also Fig.~\ref{fig:uvlf_ref}). At $z=6.3$, Euclid-deep may be able to constrain the AGN formation model by clarifying the faint end of the AGN LF at $\mathrm{M_{1450}}\sim-21$. 

In Fig.~\ref{fig:LF_Euclid}, we also plot the LFs  with the observable fraction of \citet{Merloni2014}:
\begin{equation}
\label{eq:fobs_M14}
f_{\mathrm{obs}} = 1 - 0.56 - \frac{1}{\pi} \arctan \left( \frac{43.89 - \log L_{\mathrm{X}}}{0.46}  \right),
\end{equation}
where $L_{\mathrm{X}}$ is the intrinsic X-ray luminosity in units of $\mathrm{erg \ s^{-1}}$. Since the observable fraction of \citet{Merloni2014} is higher at the bright end and lower at the faint end than that of \citet{Shirakata2019}, the LFs with \citet{Merloni2014} (black dashed lines) are larger than our fiducial model at the bright end, and the trend reverses at magnitudes $-20 \lesssim \mathrm{M_{1450}} \lesssim -19$. Fig.~\ref{fig:LF_Euclid} shows that the faint end of the LF is sensitive to the observable fraction, which can be constrained by the new surveys with Euclid.

The LSST camera on the Vera C. Rubin Observatory\footnote{\url{https://www.lsst.org/}} will discover a number of high-redshift quasars while exploring $18,000~\mathrm{deg}^2$ of the sky. In Fig.~\ref{fig:LF_LSST}, we present predictions for the AGN LF in the LSST {\it z}-band. We assume that the magnitude limit in the {\it z}-band is $26.1$, and adopt the proposed JWST area of $18,000~\mathrm{deg}^2$. As is clearly seen in this figure, the extremely large survey area and similar depth to the Euclid-deep survey will discover a large number of quasars up to $z\sim7$. LSST will clarify the shape of the AGN LF at rest-frame UV magnitudes of $\lesssim-20.5$ at $z\sim 6-7$. For quasars at $z> 7$, observations with the LSST {\it z}- and {\it y}-bands are affected by the absorption of the inter-galactic medium. Thus, to detect such high-redshift quasars, observations with near-IR filters on {\it JWST} and {\it Euclid} are effective ways. In Fig.~\ref{fig:LF_LSST}, we also plot the LFs  with the observable fraction of \citet{Merloni2014}. Like Fig.~\ref{fig:LF_Euclid}, Fig.~\ref{fig:LF_LSST} shows that the LF is sensitive to the observable fraction, which can be constrained by the new surveys with LSST, especially, at $z\sim 7$.

\section{Discussion and Summary}
\label{sec:discussion}


We have investigated the cosmic and total field-to-field variance for future galaxy and AGN surveys. To do this, we used our semi-analytic galaxy and AGN formation model, $\nu^2$GC, with a state-of-the-art cosmological {\it N}-body simulation, the Uchuu simulation. Firstly, we have shown that Uchuu-$\nu^2$GC reproduces some basic population statistics of galaxies and AGN, such as the galaxy {\it K}-band LF up to $z = 3$, the cosmic SFR density, and the AGN X-ray and UV LFs over a wide redshift range. To reproduce the latter observations, we have introduced a gas accretion time-scale onto SMBHs in our model that depends on the SMBH mass and accreted gas mass. 

Using this galaxy formation model, we have quantified the cosmic and total field-to-field variance for various survey configurations. We have shown that the cosmic variance of AGN does not significantly depend on their luminosities. This is because the typical DM halo mass in which AGN reside does not significantly depend on luminosity. In other words, this reflects the fact that the two-point correlation function of AGN is mostly luminosity independent in our model. Considering the field variance of the LFs of AGN, and in particular quasars (optically bright AGN), Poisson variance dominates the total variance.

We have forecast the apparent magnitude AGN LF and angular number density, or AGN number counts, in infrared bands that will be measured from surveys using {\it JWST}, {\it Euclid}, and LSST, taking into account the flux limit and the areas for individual surveys. From our results, the Euclid deep survey will be able to constrain the faint end of the AGN LF at $z\gtrsim 6$, even including field variance. Specifically, the Euclid deep survey can clarify the LF at the rest-frame absolute UV magnitude $\mathrm{M_{1450}}\sim-20$ at $z=6.3$. On the other hand, the Euclid wide survey will shed light on the shape of the bright end of the AGN LF. LSST will further constrain the AGN LF by its unprecedented survey area and depth. We have also shown that the faint end of the LF is sensitive to the observable fraction, implying that the faint end can be constrained to the AGN obscuration models. Finally, we predict that JWST surveys will have a difficulty constraining the AGN LF due to their small survey areas.


While we have focused on the AGN LF in this work, understanding the relations between AGN and their host galaxies is another aspect that can be achieved by our co-evolution model of galaxy and SMBHs. Using cosmological hydrodyamical simulations, mass, SFR, and structural properties of the host galaxy have been investigated (e.g. \citealt{Marshall2020}; \citealt{Habouzit2021}). These studies, however, suffer from limited AGN numbers due to their relatively small simulation volumes. We plan to examine the properties of AGN host galaxies in more detail in a future paper.

In calculating the cosmic variance in this paper, we have used the cuboidal and cubic subvolumes in a single snapshot in this paper. In other words, we have not paid attention to the redshift dependence of the AGN LF in the analysis.
This procedure can be valid if the clustering and luminosity function evolve weakly over the $\Delta z = 1$ used. As shown in Fig.~\ref{fig:xlf_ref} and \ref{fig:uvlf_ref}, there is some redshift evolution in the AGN LF over the redshift range. The effect of the redshfit evolution can be taken into account by constructing observational light cones. We defer this issue to future studies.

Although we have predicted the AGN LFs for surveys with {\it JWST}, {\it Euclid}, and LSST (Fig.~\ref{fig:LF_F200W},  \ref{fig:LF_Euclid} and \ref{fig:LF_LSST}), it is often difficult to distinguish between AGN and galaxies using only one broad-band filter. Deep imaging performed by {\it JWST} can resolve the morphology of galaxies, and AGN can be identified as point sources. Effective methods to classify these two populations are therefore needed to further derive accurate AGN LFs. We will address this in future studies.

\section*{Acknowledgements}

We are grateful to the anonymous referee for providing constructive comments. We thank Y. Matsuoka and $\nu^2$GC collaboration members for useful comments and discussion.
This research was supported by MEXT as ``Program for Promoting Researches on the Supercomputer Fugaku'' (JPMXP1020200109 and JPMXP1020230406) and JICFuS.
This study has been funded by MEXT/JSPS KAKENHI Grant Number 23K03460, 21H05449, 20K22360, 20H01950, 18H05437, JP17H04828, JP19KK0344, and JP21H01122.
Parts of this research were conducted by the Australian Research Council Centre of Excellence for All Sky Astrophysics in 3 Dimensions (ASTRO 3D), through project number CE170100013.
T.I. has been supported by IAAR Research Support Program, Chiba University, Japan. 
SAC acknowledges funding from  CONICET (PIP-2876), 
{\it Agencia Nacional de Promoci\'on de la Investigaci\'on, el Desarrollo Tecnol\'ogico y la Innovaci\'on} 
(Agencia I+D+i, PICT-2018-3743), and {\it Universidad Nacional de La Plata} (G11-150), Argentina.
RM is supported by JSPS KAKENHI Grant Number 20K14515. RM also thanks the Ministry of Science and Technology (MOST) for support through grant MOST 108-2112-M001-007-MY3, the Academia Sinica for Investigator Award AS-IA-109-M02.

We thank Instituto de Astrof\'isica de Andaluc\'ia (IAA-CSIC), Centro de Supercomputaci\'on de Galicia (CESGA) and the Spanish academic and research network (RedIRIS) in Spain for hosting 
the Uchuu products in the Skies \& Universes site (\url{http://skiesanduniverses.org/}) for cosmological simulations. 
The Uchuu simulations were carried out on Aterui II supercomputer at Center for Computational Astrophysics, CfCA, of National Astronomical Observatory of Japan, and the K computer at the RIKEN Advanced Institute for Computational Science. 
The numerical analysis was partially carried out on XC40 at the Yukawa Institute Computer Facility in Kyoto University.
The Uchuu effort has made use of the skun@IAA\_RedIRIS and skun6@IAA computer facilities managed by the IAA-CSIC in Spain.
This equipment was funded by the Spanish MICINN EU-FEDER infrastructure grant EQC2018-004366-P. 
The skun@IAA\_RedIRIS server was funded by the MICINN grant AYA2014-60641-C2-1-P. FP, AK and JR thank the support of the Spanish Ministry of Science and Innovation funding grant PGC2018-101931-B-I00. 
FP and EJ want to thank a French-Spanish international collaboration grant from CNRS and CSIC.


\section*{Data Availability}

The Uchuu DR1 products and the Uchuu-$\nu^2$GC galaxy data, a part of the Uchuu DR2 products, underlying this article are publicly available on the Skies \& Universes website at \url{http://www.skiesanduniverses.org/Simulations/Uchuu/} and \url{https://www.skiesanduniverses.org/Simulations/Uchuu/GalaxyCatalogues/} along with the necessary documentation.



\bibliographystyle{mnras}
\bibliography{ref} 

\begin{thebibliography}{}
\makeatletter
\relax
\def\mn@urlcharsother{\let\do\@makeother \do\$\do\&\do\#\do\^\do\_\do\%\do\~}
\def\mn@doi{\begingroup\mn@urlcharsother \@ifnextchar [ {\mn@doi@}
  {\mn@doi@[]}}
\def\mn@doi@[#1]#2{\def\@tempa{#1}\ifx\@tempa\@empty \href
  {http://dx.doi.org/#2} {doi:#2}\else \href {http://dx.doi.org/#2} {#1}\fi
  \endgroup}
\def\mn@eprint#1#2{\mn@eprint@#1:#2::\@nil}
\def\mn@eprint@arXiv#1{\href {http://arxiv.org/abs/#1} {{\tt arXiv:#1}}}
\def\mn@eprint@dblp#1{\href {http://dblp.uni-trier.de/rec/bibtex/#1.xml}
  {dblp:#1}}
\def\mn@eprint@#1:#2:#3:#4\@nil{\def\@tempa {#1}\def\@tempb {#2}\def\@tempc
  {#3}\ifx \@tempc \@empty \let \@tempc \@tempb \let \@tempb \@tempa \fi \ifx
  \@tempb \@empty \def\@tempb {arXiv}\fi \@ifundefined
  {mn@eprint@\@tempb}{\@tempb:\@tempc}{\expandafter \expandafter \csname
  mn@eprint@\@tempb\endcsname \expandafter{\@tempc}}}

\bibitem[\protect\citeauthoryear{{Aird}, {Coil}, {Georgakakis}, {Nandra},
  {Barro}  \& {P{\'e}rez-Gonz{\'a}lez}}{{Aird} et~al.}{2015}]{Aird2015}
{Aird} J.,  {Coil} A.~L.,  {Georgakakis} A.,  {Nandra} K.,  {Barro} G.,
  {P{\'e}rez-Gonz{\'a}lez} P.~G.,  2015, \mn@doi [\mnras]
  {10.1093/mnras/stv1062}, \href
  {https://ui.adsabs.harvard.edu/abs/2015MNRAS.451.1892A} {451, 1892}

\bibitem[\protect\citeauthoryear{{Akiyama} et~al.,}{{Akiyama}
  et~al.}{2018}]{Akiyama2018}
{Akiyama} M.,  et~al., 2018, \mn@doi [\pasj] {10.1093/pasj/psx091}, \href
  {https://ui.adsabs.harvard.edu/abs/2018PASJ...70S..34A} {70, S34}

\bibitem[\protect\citeauthoryear{{Amarantidis} et~al.,}{{Amarantidis}
  et~al.}{2019}]{Amarantidis2019}
{Amarantidis} S.,  et~al., 2019, \mn@doi [\mnras] {10.1093/mnras/stz551}, \href
  {https://ui.adsabs.harvard.edu/abs/2019MNRAS.485.2694A} {485, 2694}

\bibitem[\protect\citeauthoryear{{Atek}, {Richard}, {Kneib}  \&
  {Schaerer}}{{Atek} et~al.}{2018}]{Atek2018}
{Atek} H.,  {Richard} J.,  {Kneib} J.-P.,   {Schaerer} D.,  2018, \mn@doi
  [\mnras] {10.1093/mnras/sty1820}, \href
  {https://ui.adsabs.harvard.edu/abs/2018MNRAS.479.5184A} {479, 5184}

\bibitem[\protect\citeauthoryear{{Behroozi}, {Wechsler}  \& {Wu}}{{Behroozi}
  et~al.}{2013a}]{Behroozi2013a}
{Behroozi} P.~S.,  {Wechsler} R.~H.,   {Wu} H.-Y.,  2013a, \mn@doi [\apj]
  {10.1088/0004-637X/762/2/109}, \href
  {https://ui.adsabs.harvard.edu/abs/2013ApJ...762..109B} {762, 109}

\bibitem[\protect\citeauthoryear{{Behroozi}, {Wechsler}, {Wu}, {Busha},
  {Klypin}  \& {Primack}}{{Behroozi} et~al.}{2013b}]{Behroozi2013b}
{Behroozi} P.~S.,  {Wechsler} R.~H.,  {Wu} H.-Y.,  {Busha} M.~T.,  {Klypin}
  A.~A.,   {Primack} J.~R.,  2013b, \mn@doi [\apj]
  {10.1088/0004-637X/763/1/18}, \href
  {https://ui.adsabs.harvard.edu/abs/2013ApJ...763...18B} {763, 18}

\bibitem[\protect\citeauthoryear{{Behroozi}, {Wechsler}  \&
  {Conroy}}{{Behroozi} et~al.}{2013c}]{Behroozi2013c}
{Behroozi} P.~S.,  {Wechsler} R.~H.,   {Conroy} C.,  2013c, \mn@doi [\apj]
  {10.1088/0004-637X/770/1/57}, \href
  {https://ui.adsabs.harvard.edu/abs/2013ApJ...770...57B} {770, 57}

\bibitem[\protect\citeauthoryear{{Bhowmick}, {Somerville}, {Di Matteo},
  {Wilkins}, {Feng}  \& {Tenneti}}{{Bhowmick} et~al.}{2020}]{Bhowmick2020}
{Bhowmick} A.~K.,  {Somerville} R.~S.,  {Di Matteo} T.,  {Wilkins} S.,  {Feng}
  Y.,   {Tenneti} A.,  2020, \mn@doi [\mnras] {10.1093/mnras/staa1605}, \href
  {https://ui.adsabs.harvard.edu/abs/2020MNRAS.496..754B} {496, 754}

\bibitem[\protect\citeauthoryear{{Bongiorno} et~al.,}{{Bongiorno}
  et~al.}{2007}]{Bongiorno2007}
{Bongiorno} A.,  et~al., 2007, \mn@doi [\aap] {10.1051/0004-6361:20077611},
  \href {https://ui.adsabs.harvard.edu/abs/2007A&A...472..443B} {472, 443}

\bibitem[\protect\citeauthoryear{{Bouwens} et~al.,}{{Bouwens}
  et~al.}{2015}]{Bouwens2015}
{Bouwens} R.~J.,  et~al., 2015, \mn@doi [\apj] {10.1088/0004-637X/803/1/34},
  \href {https://ui.adsabs.harvard.edu/abs/2015ApJ...803...34B} {803, 34}

\bibitem[\protect\citeauthoryear{{Brusa} et~al.,}{{Brusa}
  et~al.}{2010}]{Brusa2010}
{Brusa} M.,  et~al., 2010, \mn@doi [\apj] {10.1088/0004-637X/716/1/348}, \href
  {https://ui.adsabs.harvard.edu/abs/2010ApJ...716..348B} {716, 348}

\bibitem[\protect\citeauthoryear{{Caputi}, {McLure}, {Dunlop}, {Cirasuolo}  \&
  {Schael}}{{Caputi} et~al.}{2006}]{Caputi2006}
{Caputi} K.~I.,  {McLure} R.~J.,  {Dunlop} J.~S.,  {Cirasuolo} M.,   {Schael}
  A.~M.,  2006, \mn@doi [\mnras] {10.1111/j.1365-2966.2005.09887.x}, \href
  {https://ui.adsabs.harvard.edu/abs/2006MNRAS.366..609C} {366, 609}

\bibitem[\protect\citeauthoryear{{Cirasuolo}, {McLure}, {Dunlop}, {Almaini},
  {Foucaud}  \& {Simpson}}{{Cirasuolo} et~al.}{2010}]{Cirasuolo2010}
{Cirasuolo} M.,  {McLure} R.~J.,  {Dunlop} J.~S.,  {Almaini} O.,  {Foucaud} S.,
    {Simpson} C.,  2010, \mn@doi [\mnras] {10.1111/j.1365-2966.2009.15710.x},
  \href {https://ui.adsabs.harvard.edu/abs/2010MNRAS.401.1166C} {401, 1166}

\bibitem[\protect\citeauthoryear{{Civano} et~al.,}{{Civano}
  et~al.}{2011}]{Civano2011}
{Civano} F.,  et~al., 2011, \mn@doi [\apj] {10.1088/0004-637X/741/2/91}, \href
  {https://ui.adsabs.harvard.edu/abs/2011ApJ...741...91C} {741, 91}

\bibitem[\protect\citeauthoryear{{Croom} et~al.,}{{Croom}
  et~al.}{2005}]{Croom2005}
{Croom} S.~M.,  et~al., 2005, \mnras, 356, 415

\bibitem[\protect\citeauthoryear{{Croom} et~al.,}{{Croom}
  et~al.}{2009}]{Croom2009}
{Croom} S.~M.,  et~al., 2009, \mn@doi [\mnras]
  {10.1111/j.1365-2966.2009.15398.x}, \href
  {https://ui.adsabs.harvard.edu/abs/2009MNRAS.399.1755C} {399, 1755}

\bibitem[\protect\citeauthoryear{{Driver} et~al.,}{{Driver}
  et~al.}{2012}]{Driver2012}
{Driver} S.~P.,  et~al., 2012, \mn@doi [\mnras]
  {10.1111/j.1365-2966.2012.22036.x}, \href
  {https://ui.adsabs.harvard.edu/abs/2012MNRAS.427.3244D} {427, 3244}

\bibitem[\protect\citeauthoryear{{Driver} et~al.,}{{Driver}
  et~al.}{2018}]{Driver2018}
{Driver} S.~P.,  et~al., 2018, \mn@doi [\mnras] {10.1093/mnras/stx2728}, \href
  {https://ui.adsabs.harvard.edu/abs/2018MNRAS.475.2891D} {475, 2891}

\bibitem[\protect\citeauthoryear{{Duras} et~al.,}{{Duras}
  et~al.}{2020}]{Duras2020}
{Duras} F.,  et~al., 2020, \mn@doi [\aap] {10.1051/0004-6361/201936817}, \href
  {https://ui.adsabs.harvard.edu/abs/2020A&A...636A..73D} {636, A73}

\bibitem[\protect\citeauthoryear{{Enoki}, {Ishiyama}, {Kobayashi}  \&
  {Nagashima}}{{Enoki} et~al.}{2014}]{Enoki2014}
{Enoki} M.,  {Ishiyama} T.,  {Kobayashi} M. A.~R.,   {Nagashima} M.,  2014,
  \mn@doi [\apj] {10.1088/0004-637X/794/1/69}, \href
  {https://ui.adsabs.harvard.edu/abs/2014ApJ...794...69E} {794, 69}

\bibitem[\protect\citeauthoryear{{Fanidakis} et~al.,}{{Fanidakis}
  et~al.}{2012}]{Fanidakis2012}
{Fanidakis} N.,  et~al., 2012, \mn@doi [\mnras]
  {10.1111/j.1365-2966.2011.19931.x}, \href
  {https://ui.adsabs.harvard.edu/abs/2012MNRAS.419.2797F} {419, 2797}

\bibitem[\protect\citeauthoryear{{Feng}, {Di-Matteo}, {Croft}, {Bird},
  {Battaglia}  \& {Wilkins}}{{Feng} et~al.}{2016}]{Feng2016}
{Feng} Y.,  {Di-Matteo} T.,  {Croft} R.~A.,  {Bird} S.,  {Battaglia} N.,
  {Wilkins} S.,  2016, \mn@doi [\mnras] {10.1093/mnras/stv2484}, \href
  {https://ui.adsabs.harvard.edu/abs/2016MNRAS.455.2778F} {455, 2778}

\bibitem[\protect\citeauthoryear{{Fiore} et~al.,}{{Fiore}
  et~al.}{2012}]{Fiore2012}
{Fiore} F.,  et~al., 2012, \mn@doi [\aap] {10.1051/0004-6361/201117581}, \href
  {https://ui.adsabs.harvard.edu/abs/2012A&A...537A..16F} {537, A16}

\bibitem[\protect\citeauthoryear{{Fontanot}, {Cristiani}, {Monaco}, {Nonino},
  {Vanzella}, {Brandt}, {Grazian}  \& {Mao}}{{Fontanot}
  et~al.}{2007}]{Fontanot2007}
{Fontanot} F.,  {Cristiani} S.,  {Monaco} P.,  {Nonino} M.,  {Vanzella} E.,
  {Brandt} W.~N.,  {Grazian} A.,   {Mao} J.,  2007, \mn@doi [\aap]
  {10.1051/0004-6361:20066073}, \href
  {https://ui.adsabs.harvard.edu/abs/2007A&A...461...39F} {461, 39}

\bibitem[\protect\citeauthoryear{{Gardner} et~al.,}{{Gardner}
  et~al.}{2006}]{Gardner2006}
{Gardner} J.~P.,  et~al., 2006, \mn@doi [\ssr] {10.1007/s11214-006-8315-7},
  \href {https://ui.adsabs.harvard.edu/abs/2006SSRv..123..485G} {123, 485}

\bibitem[\protect\citeauthoryear{{Giallongo} et~al.,}{{Giallongo}
  et~al.}{2019}]{Giallongo2019}
{Giallongo} E.,  et~al., 2019, \mn@doi [\apj] {10.3847/1538-4357/ab39e1}, \href
  {https://ui.adsabs.harvard.edu/abs/2019ApJ...884...19G} {884, 19}

\bibitem[\protect\citeauthoryear{{Grazian} et~al.,}{{Grazian}
  et~al.}{2020}]{Grazian2020}
{Grazian} A.,  et~al., 2020, \mn@doi [\apj] {10.3847/1538-4357/ab99a3}, \href
  {https://ui.adsabs.harvard.edu/abs/2020ApJ...897...94G} {897, 94}

\bibitem[\protect\citeauthoryear{{Griffin}, {Lacey}, {Gonzalez-Perez}, {Lagos},
  {Baugh}  \& {Fanidakis}}{{Griffin} et~al.}{2019}]{Griffin2019}
{Griffin} A.~J.,  {Lacey} C.~G.,  {Gonzalez-Perez} V.,  {Lagos} C. d.~P.,
  {Baugh} C.~M.,   {Fanidakis} N.,  2019, \mn@doi [\mnras]
  {10.1093/mnras/stz1216}, \href
  {https://ui.adsabs.harvard.edu/abs/2019MNRAS.487..198G} {487, 198}

\bibitem[\protect\citeauthoryear{{Griffin}, {Lacey}, {Gonzalez-Perez}, {Lagos},
  {Baugh}  \& {Fanidakis}}{{Griffin} et~al.}{2020}]{Griffin2020}
{Griffin} A.~J.,  {Lacey} C.~G.,  {Gonzalez-Perez} V.,  {Lagos} C. d.~P.,
  {Baugh} C.~M.,   {Fanidakis} N.,  2020, \mn@doi [\mnras]
  {10.1093/mnras/staa024}, \href
  {https://ui.adsabs.harvard.edu/abs/2020MNRAS.492.2535G} {492, 2535}

\bibitem[\protect\citeauthoryear{{Habouzit}, {Volonteri}, {Somerville},
  {Dubois}, {Peirani}, {Pichon}  \& {Devriendt}}{{Habouzit}
  et~al.}{2019}]{Habouzit2019}
{Habouzit} M.,  {Volonteri} M.,  {Somerville} R.~S.,  {Dubois} Y.,  {Peirani}
  S.,  {Pichon} C.,   {Devriendt} J.,  2019, \mn@doi [\mnras]
  {10.1093/mnras/stz2105}, \href
  {https://ui.adsabs.harvard.edu/abs/2019MNRAS.489.1206H} {489, 1206}

\bibitem[\protect\citeauthoryear{{Habouzit} et~al.,}{{Habouzit}
  et~al.}{2021}]{Habouzit2021}
{Habouzit} M.,  et~al., 2021, \mn@doi [\mnras] {10.1093/mnras/stab496}, \href
  {https://ui.adsabs.harvard.edu/abs/2021MNRAS.503.1940H} {503, 1940}

\bibitem[\protect\citeauthoryear{{He} et~al.,}{{He} et~al.}{2018}]{He2018}
{He} W.,  et~al., 2018, \mn@doi [\pasj] {10.1093/pasj/psx129}, \href
  {https://ui.adsabs.harvard.edu/abs/2018PASJ...70S..33H} {70, S33}

\bibitem[\protect\citeauthoryear{{Henriques}, {Maraston}, {Monaco}, {Fontanot},
  {Menci}, {De Lucia}  \& {Tonini}}{{Henriques} et~al.}{2011}]{Henriques2011}
{Henriques} B.,  {Maraston} C.,  {Monaco} P.,  {Fontanot} F.,  {Menci} N.,  {De
  Lucia} G.,   {Tonini} C.,  2011, \mn@doi [\mnras]
  {10.1111/j.1365-2966.2011.18972.x}, \href
  {https://ui.adsabs.harvard.edu/abs/2011MNRAS.415.3571H} {415, 3571}

\bibitem[\protect\citeauthoryear{{Hirschmann}, {Somerville}, {Naab}  \&
  {Burkert}}{{Hirschmann} et~al.}{2012}]{Hirschmann2012}
{Hirschmann} M.,  {Somerville} R.~S.,  {Naab} T.,   {Burkert} A.,  2012,
  \mn@doi [\mnras] {10.1111/j.1365-2966.2012.21626.x}, \href
  {https://ui.adsabs.harvard.edu/abs/2012MNRAS.426..237H} {426, 237}

\bibitem[\protect\citeauthoryear{{Hopkins}}{{Hopkins}}{2004}]{Hopkins2004}
{Hopkins} A.~M.,  2004, \mn@doi [\apj] {10.1086/424032}, \href
  {https://ui.adsabs.harvard.edu/abs/2004ApJ...615..209H} {615, 209}

\bibitem[\protect\citeauthoryear{{Hopkins}, {Richards}  \&
  {Hernquist}}{{Hopkins} et~al.}{2007}]{Hopkins2007}
{Hopkins} P.~F.,  {Richards} G.~T.,   {Hernquist} L.,  2007, \mn@doi [\apj]
  {10.1086/509629}, \href
  {https://ui.adsabs.harvard.edu/abs/2007ApJ...654..731H} {654, 731}

\bibitem[\protect\citeauthoryear{{Hopkins}, {Cox}, {Younger}  \&
  {Hernquist}}{{Hopkins} et~al.}{2009}]{Hopkins2009ApJ...691.1168H}
{Hopkins} P.~F.,  {Cox} T.~J.,  {Younger} J.~D.,   {Hernquist} L.,  2009,
  \mn@doi [\apj] {10.1088/0004-637X/691/2/1168}, \href
  {https://ui.adsabs.harvard.edu/abs/2009ApJ...691.1168H} {691, 1168}

\bibitem[\protect\citeauthoryear{{Ishigaki}, {Kawamata}, {Ouchi}, {Oguri},
  {Shimasaku}  \& {Ono}}{{Ishigaki} et~al.}{2018}]{Ishigaki2018}
{Ishigaki} M.,  {Kawamata} R.,  {Ouchi} M.,  {Oguri} M.,  {Shimasaku} K.,
  {Ono} Y.,  2018, \mn@doi [\apj] {10.3847/1538-4357/aaa544}, \href
  {https://ui.adsabs.harvard.edu/abs/2018ApJ...854...73I} {854, 73}

\bibitem[\protect\citeauthoryear{{Ishiyama}, {Fukushige}  \&
  {Makino}}{{Ishiyama} et~al.}{2009}]{Ishiyama2009}
{Ishiyama} T.,  {Fukushige} T.,   {Makino} J.,  2009, \mn@doi [\pasj]
  {10.1093/pasj/61.6.1319}, \href
  {https://ui.adsabs.harvard.edu/abs/2009PASJ...61.1319I} {61, 1319}

\bibitem[\protect\citeauthoryear{{Ishiyama}, {Nitadori}  \&
  {Makino}}{{Ishiyama} et~al.}{2012}]{Ishiyama2012}
{Ishiyama} T.,  {Nitadori} K.,   {Makino} J.,  2012, arXiv e-prints, \href
  {https://ui.adsabs.harvard.edu/abs/2012arXiv1211.4406I} {p. arXiv:1211.4406}

\bibitem[\protect\citeauthoryear{{Ishiyama}, {Enoki}, {Kobayashi}, {Makiya},
  {Nagashima}  \& {Oogi}}{{Ishiyama} et~al.}{2015}]{Ishiyama2015}
{Ishiyama} T.,  {Enoki} M.,  {Kobayashi} M.~A.~R.,  {Makiya} R.,  {Nagashima}
  M.,   {Oogi} T.,  2015, \pasj, 67, 61

\bibitem[\protect\citeauthoryear{{Ishiyama} et~al.,}{{Ishiyama}
  et~al.}{2021}]{Ishiyama2021}
{Ishiyama} T.,  et~al., 2021, \mn@doi [\mnras] {10.1093/mnras/stab1755}, \href
  {https://ui.adsabs.harvard.edu/abs/2021MNRAS.506.4210I} {506, 4210}

\bibitem[\protect\citeauthoryear{{Kawaguchi}}{{Kawaguchi}}{2003}]{Kawaguchi2003}
{Kawaguchi} T.,  2003, \mn@doi [\apj] {10.1086/376404}, \href
  {https://ui.adsabs.harvard.edu/abs/2003ApJ...593...69K} {593, 69}

\bibitem[\protect\citeauthoryear{{Kawaguchi}, {Shimura}  \&
  {Mineshige}}{{Kawaguchi} et~al.}{2001}]{Kawaguchi2001}
{Kawaguchi} T.,  {Shimura} T.,   {Mineshige} S.,  2001, \mn@doi [\apj]
  {10.1086/318297}, \href
  {https://ui.adsabs.harvard.edu/abs/2001ApJ...546..966K} {546, 966}

\bibitem[\protect\citeauthoryear{{Krolewski} \& {Eisenstein}}{{Krolewski} \&
  {Eisenstein}}{2015}]{Krolewski2015}
{Krolewski} A.~G.,  {Eisenstein} D.~J.,  2015, \apj, 803, 4

\bibitem[\protect\citeauthoryear{{LSST Science Collaboration} et~al.,}{{LSST
  Science Collaboration} et~al.}{2009}]{LSST2009}
{LSST Science Collaboration} et~al., 2009, arXiv e-prints, \href
  {https://ui.adsabs.harvard.edu/abs/2009arXiv0912.0201L} {p. arXiv:0912.0201}

\bibitem[\protect\citeauthoryear{{Lacey} et~al.,}{{Lacey}
  et~al.}{2016}]{Lacey2016}
{Lacey} C.~G.,  et~al., 2016, \mn@doi [\mnras] {10.1093/mnras/stw1888}, \href
  {https://ui.adsabs.harvard.edu/abs/2016MNRAS.462.3854L} {462, 3854}

\bibitem[\protect\citeauthoryear{{Lagos} et~al.,}{{Lagos}
  et~al.}{2019}]{Lagos2019}
{Lagos} C. d.~P.,  et~al., 2019, \mn@doi [\mnras] {10.1093/mnras/stz2427},
  \href {https://ui.adsabs.harvard.edu/abs/2019MNRAS.489.4196L} {489, 4196}

\bibitem[\protect\citeauthoryear{{Laureijs} et~al.,}{{Laureijs}
  et~al.}{2011}]{Euclid2011}
{Laureijs} R.,  et~al., 2011, arXiv e-prints, \href
  {https://ui.adsabs.harvard.edu/abs/2011arXiv1110.3193L} {p. arXiv:1110.3193}

\bibitem[\protect\citeauthoryear{{Livermore}, {Finkelstein}  \&
  {Lotz}}{{Livermore} et~al.}{2017}]{Livermore2017}
{Livermore} R.~C.,  {Finkelstein} S.~L.,   {Lotz} J.~M.,  2017, \mn@doi [\apj]
  {10.3847/1538-4357/835/2/113}, \href
  {https://ui.adsabs.harvard.edu/abs/2017ApJ...835..113L} {835, 113}

\bibitem[\protect\citeauthoryear{{Makiya} et~al.,}{{Makiya}
  et~al.}{2016}]{Makiya2016}
{Makiya} R.,  et~al., 2016, \mn@doi [\pasj] {10.1093/pasj/psw005}, \href
  {https://ui.adsabs.harvard.edu/abs/2016PASJ...68...25M} {68, 25}

\bibitem[\protect\citeauthoryear{{Marconi}, {Risaliti}, {Gilli}, {Hunt},
  {Maiolino}  \& {Salvati}}{{Marconi} et~al.}{2004}]{Marconi2004}
{Marconi} A.,  {Risaliti} G.,  {Gilli} R.,  {Hunt} L.~K.,  {Maiolino} R.,
  {Salvati} M.,  2004, \mnras, 351, 169

\bibitem[\protect\citeauthoryear{{Marshall}, {Ni}, {Di Matteo}, {Wyithe},
  {Wilkins}, {Croft}  \& {Kuusisto}}{{Marshall} et~al.}{2020}]{Marshall2020}
{Marshall} M.~A.,  {Ni} Y.,  {Di Matteo} T.,  {Wyithe} J. S.~B.,  {Wilkins} S.,
   {Croft} R. A.~C.,   {Kuusisto} J.~K.,  2020, \mn@doi [\mnras]
  {10.1093/mnras/staa2982}, \href
  {https://ui.adsabs.harvard.edu/abs/2020MNRAS.499.3819M} {499, 3819}

\bibitem[\protect\citeauthoryear{{Marshall}, {Wyithe}, {Windhorst}, {Matteo},
  {Ni}, {Wilkins}, {Croft}  \& {Mechtley}}{{Marshall}
  et~al.}{2021}]{Marshall2021}
{Marshall} M.~A.,  {Wyithe} J. S.~B.,  {Windhorst} R.~A.,  {Matteo} T.~D.,
  {Ni} Y.,  {Wilkins} S.,  {Croft} R. A.~C.,   {Mechtley} M.,  2021, \mn@doi
  [\mnras] {10.1093/mnras/stab1763}, \href
  {https://ui.adsabs.harvard.edu/abs/2021MNRAS.506.1209M} {506, 1209}

\bibitem[\protect\citeauthoryear{{Matsuoka} et~al.,}{{Matsuoka}
  et~al.}{2018}]{Matsuoka2018}
{Matsuoka} Y.,  et~al., 2018, \mn@doi [\apj] {10.3847/1538-4357/aaee7a}, \href
  {https://ui.adsabs.harvard.edu/abs/2018ApJ...869..150M} {869, 150}

\bibitem[\protect\citeauthoryear{{McConnell} \& {Ma}}{{McConnell} \&
  {Ma}}{2013}]{McConnell2013}
{McConnell} N.~J.,  {Ma} C.-P.,  2013, \apj, 764, 184

\bibitem[\protect\citeauthoryear{{Merloni} et~al.,}{{Merloni}
  et~al.}{2014}]{Merloni2014}
{Merloni} A.,  et~al., 2014, \mn@doi [\mnras] {10.1093/mnras/stt2149}, \href
  {https://ui.adsabs.harvard.edu/abs/2014MNRAS.437.3550M} {437, 3550}

\bibitem[\protect\citeauthoryear{{Mineshige}, {Kawaguchi}, {Takeuchi}  \&
  {Hayashida}}{{Mineshige} et~al.}{2000}]{Mineshige2000}
{Mineshige} S.,  {Kawaguchi} T.,  {Takeuchi} M.,   {Hayashida} K.,  2000,
  \pasj, 52, 499

\bibitem[\protect\citeauthoryear{{Moster}, {Somerville}, {Newman}  \&
  {Rix}}{{Moster} et~al.}{2011}]{Moster2011}
{Moster} B.~P.,  {Somerville} R.~S.,  {Newman} J.~A.,   {Rix} H.-W.,  2011,
  \mn@doi [\apj] {10.1088/0004-637X/731/2/113}, \href
  {https://ui.adsabs.harvard.edu/abs/2011ApJ...731..113M} {731, 113}

\bibitem[\protect\citeauthoryear{{Myers}, {Brunner}, {Nichol}, {Richards},
  {Schneider}  \& {Bahcall}}{{Myers} et~al.}{2007}]{Myers2007}
{Myers} A.~D.,  {Brunner} R.~J.,  {Nichol} R.~C.,  {Richards} G.~T.,
  {Schneider} D.~P.,   {Bahcall} N.~A.,  2007, \apj, 658, 85

\bibitem[\protect\citeauthoryear{{Nagashima}, {Yahagi}, {Enoki}, {Yoshii}  \&
  {Gouda}}{{Nagashima} et~al.}{2005}]{Nagashima2005}
{Nagashima} M.,  {Yahagi} H.,  {Enoki} M.,  {Yoshii} Y.,   {Gouda} N.,  2005,
  \apj, 634, 26

\bibitem[\protect\citeauthoryear{{Ni}, {Di Matteo}  \& {Feng}}{{Ni}
  et~al.}{2020a}]{Ni2020b}
{Ni} Y.,  {Di Matteo} T.,   {Feng} Y.,  2020a, arXiv e-prints, \href
  {https://ui.adsabs.harvard.edu/abs/2020arXiv201204714N} {p. arXiv:2012.04714}

\bibitem[\protect\citeauthoryear{{Ni}, {Di Matteo}, {Gilli}, {Croft}, {Feng}
  \& {Norman}}{{Ni} et~al.}{2020b}]{Ni2020}
{Ni} Y.,  {Di Matteo} T.,  {Gilli} R.,  {Croft} R. A.~C.,  {Feng} Y.,
  {Norman} C.,  2020b, \mn@doi [\mnras] {10.1093/mnras/staa1313}, \href
  {https://ui.adsabs.harvard.edu/abs/2020MNRAS.495.2135N} {495, 2135}

\bibitem[\protect\citeauthoryear{{Niida} et~al.,}{{Niida}
  et~al.}{2020}]{Niida2020}
{Niida} M.,  et~al., 2020, \mn@doi [\apj] {10.3847/1538-4357/abbe11}, \href
  {https://ui.adsabs.harvard.edu/abs/2020ApJ...904...89N} {904, 89}

\bibitem[\protect\citeauthoryear{{Ogura} et~al.,}{{Ogura}
  et~al.}{2020}]{Ogura2020}
{Ogura} K.,  et~al., 2020, \mn@doi [\apj] {10.3847/1538-4357/ab8631}, \href
  {https://ui.adsabs.harvard.edu/abs/2020ApJ...895....9O} {895, 9}

\bibitem[\protect\citeauthoryear{{Ohsuga}, {Mori}, {Nakamoto}  \&
  {Mineshige}}{{Ohsuga} et~al.}{2005}]{Ohsuga2005}
{Ohsuga} K.,  {Mori} M.,  {Nakamoto} T.,   {Mineshige} S.,  2005, \apj, 628,
  368

\bibitem[\protect\citeauthoryear{{Oogi}, {Enoki}, {Ishiyama}, {Kobayashi},
  {Makiya}  \& {Nagashima}}{{Oogi} et~al.}{2016}]{Oogi2016}
{Oogi} T.,  {Enoki} M.,  {Ishiyama} T.,  {Kobayashi} M. A.~R.,  {Makiya} R.,
  {Nagashima} M.,  2016, \mn@doi [\mnras] {10.1093/mnrasl/slv169}, \href
  {https://ui.adsabs.harvard.edu/abs/2016MNRAS.456L..30O} {456, L30}

\bibitem[\protect\citeauthoryear{{Oogi}, {Enoki}, {Ishiyama}, {Kobayashi},
  {Makiya}, {Nagashima}, {Okamoto}  \& {Shirakata}}{{Oogi}
  et~al.}{2017}]{Oogi2017}
{Oogi} T.,  {Enoki} M.,  {Ishiyama} T.,  {Kobayashi} M. A.~R.,  {Makiya} R.,
  {Nagashima} M.,  {Okamoto} T.,   {Shirakata} H.,  2017, \mn@doi [\mnras]
  {10.1093/mnrasl/slx102}, \href
  {https://ui.adsabs.harvard.edu/abs/2017MNRAS.471L..21O} {471, L21}

\bibitem[\protect\citeauthoryear{{Oogi}, {Shirakata}, {Nagashima},
  {Nishimichi}, {Kawaguchi}, {Okamoto}, {Ishiyama}  \& {Enoki}}{{Oogi}
  et~al.}{2020}]{Oogi2020}
{Oogi} T.,  {Shirakata} H.,  {Nagashima} M.,  {Nishimichi} T.,  {Kawaguchi} T.,
   {Okamoto} T.,  {Ishiyama} T.,   {Enoki} M.,  2020, \mn@doi [\mnras]
  {10.1093/mnras/staa1961}, \href
  {https://ui.adsabs.harvard.edu/abs/2020MNRAS.497....1O} {497, 1}

\bibitem[\protect\citeauthoryear{{Padmanabhan}, {White}, {Norberg}  \&
  {Porciani}}{{Padmanabhan} et~al.}{2009}]{Padmanabhan2009}
{Padmanabhan} N.,  {White} M.,  {Norberg} P.,   {Porciani} C.,  2009, \mnras,
  397, 1862

\bibitem[\protect\citeauthoryear{{Piana}, {Dayal}, {Volonteri}  \&
  {Choudhury}}{{Piana} et~al.}{2021}]{Piana2021}
{Piana} O.,  {Dayal} P.,  {Volonteri} M.,   {Choudhury} T.~R.,  2021, \mn@doi
  [\mnras] {10.1093/mnras/staa3363}, \href
  {https://ui.adsabs.harvard.edu/abs/2021MNRAS.500.2146P} {500, 2146}

\bibitem[\protect\citeauthoryear{{Planck Collaboration} et~al.,}{{Planck
  Collaboration} et~al.}{2020}]{Planck2020}
{Planck Collaboration} et~al., 2020, \mn@doi [\aap]
  {10.1051/0004-6361/201833910}, \href
  {https://ui.adsabs.harvard.edu/abs/2020A&A...641A...6P} {641, A6}

\bibitem[\protect\citeauthoryear{{Ricarte} \& {Natarajan}}{{Ricarte} \&
  {Natarajan}}{2018}]{Ricarte2018}
{Ricarte} A.,  {Natarajan} P.,  2018, \mn@doi [\mnras] {10.1093/mnras/sty2448},
  \href {https://ui.adsabs.harvard.edu/abs/2018MNRAS.481.3278R} {481, 3278}

\bibitem[\protect\citeauthoryear{{Richards} et~al.,}{{Richards}
  et~al.}{2005}]{Richards2005}
{Richards} G.~T.,  et~al., 2005, \mn@doi [\mnras]
  {10.1111/j.1365-2966.2005.09096.x}, \href
  {https://ui.adsabs.harvard.edu/abs/2005MNRAS.360..839R} {360, 839}

\bibitem[\protect\citeauthoryear{{Ross} et~al.,}{{Ross}
  et~al.}{2009}]{Ross2009}
{Ross} N.~P.,  et~al., 2009, \apj, 697, 1634

\bibitem[\protect\citeauthoryear{{Saracco} et~al.,}{{Saracco}
  et~al.}{2006}]{Saracco2006}
{Saracco} P.,  et~al., 2006, \mn@doi [\mnras]
  {10.1111/j.1365-2966.2006.09967.x}, \href
  {https://ui.adsabs.harvard.edu/abs/2006MNRAS.367..349S} {367, 349}

\bibitem[\protect\citeauthoryear{{Shen} et~al.,}{{Shen}
  et~al.}{2007}]{Shen2007}
{Shen} Y.,  et~al., 2007, \aj, 133, 2222

\bibitem[\protect\citeauthoryear{{Shirakata} et~al.,}{{Shirakata}
  et~al.}{2016}]{Shirakata2016}
{Shirakata} H.,  et~al., 2016, \mn@doi [\mnras] {10.1093/mnras/stw1798}, \href
  {https://ui.adsabs.harvard.edu/abs/2016MNRAS.461.4389S} {461, 4389}

\bibitem[\protect\citeauthoryear{{Shirakata} et~al.,}{{Shirakata}
  et~al.}{2019}]{Shirakata2019}
{Shirakata} H.,  et~al., 2019, \mn@doi [\mnras] {10.1093/mnras/sty2958}, \href
  {https://ui.adsabs.harvard.edu/abs/2019MNRAS.482.4846S} {482, 4846}

\bibitem[\protect\citeauthoryear{{Siana} et~al.,}{{Siana}
  et~al.}{2008}]{Siana2008}
{Siana} B.,  et~al., 2008, \mn@doi [\apj] {10.1086/527025}, \href
  {https://ui.adsabs.harvard.edu/abs/2008ApJ...675...49S} {675, 49}

\bibitem[\protect\citeauthoryear{{Somerville} \& {Dav{\'e}}}{{Somerville} \&
  {Dav{\'e}}}{2015}]{Somerville_Dave2015}
{Somerville} R.~S.,  {Dav{\'e}} R.,  2015, \mn@doi [\araa]
  {10.1146/annurev-astro-082812-140951}, \href
  {https://ui.adsabs.harvard.edu/abs/2015ARA&A..53...51S} {53, 51}

\bibitem[\protect\citeauthoryear{{Somerville}, {Lee}, {Ferguson}, {Gardner},
  {Moustakas}  \& {Giavalisco}}{{Somerville} et~al.}{2004}]{Somerville2004}
{Somerville} R.~S.,  {Lee} K.,  {Ferguson} H.~C.,  {Gardner} J.~P.,
  {Moustakas} L.~A.,   {Giavalisco} M.,  2004, \mn@doi [\apjl]
  {10.1086/378628}, \href
  {https://ui.adsabs.harvard.edu/abs/2004ApJ...600L.171S} {600, L171}

\bibitem[\protect\citeauthoryear{{Somerville}, {Gilmore}, {Primack}  \&
  {Dom{\'\i}nguez}}{{Somerville} et~al.}{2012}]{Somerville2012}
{Somerville} R.~S.,  {Gilmore} R.~C.,  {Primack} J.~R.,   {Dom{\'\i}nguez} A.,
  2012, \mn@doi [\mnras] {10.1111/j.1365-2966.2012.20490.x}, \href
  {https://ui.adsabs.harvard.edu/abs/2012MNRAS.423.1992S} {423, 1992}

\bibitem[\protect\citeauthoryear{{Trenti} \& {Stiavelli}}{{Trenti} \&
  {Stiavelli}}{2008}]{Trenti2008}
{Trenti} M.,  {Stiavelli} M.,  2008, \mn@doi [\apj] {10.1086/528674}, \href
  {https://ui.adsabs.harvard.edu/abs/2008ApJ...676..767T} {676, 767}

\bibitem[\protect\citeauthoryear{{Ucci} et~al.,}{{Ucci}
  et~al.}{2021}]{Ucci2021}
{Ucci} G.,  et~al., 2021, \mn@doi [\mnras] {10.1093/mnras/stab1229}, \href
  {https://ui.adsabs.harvard.edu/abs/2021MNRAS.506..202U} {506, 202}

\bibitem[\protect\citeauthoryear{{Ueda}, {Akiyama}, {Hasinger}, {Miyaji}  \&
  {Watson}}{{Ueda} et~al.}{2014}]{Ueda2014}
{Ueda} Y.,  {Akiyama} M.,  {Hasinger} G.,  {Miyaji} T.,   {Watson} M.~G.,
  2014, \mn@doi [\apj] {10.1088/0004-637X/786/2/104}, \href
  {https://ui.adsabs.harvard.edu/abs/2014ApJ...786..104U} {786, 104}

\bibitem[\protect\citeauthoryear{{Wang} et~al.,}{{Wang}
  et~al.}{2019}]{Wang2019}
{Wang} F.,  et~al., 2019, \mn@doi [\apj] {10.3847/1538-4357/ab2be5}, \href
  {https://ui.adsabs.harvard.edu/abs/2019ApJ...884...30W} {884, 30}

\bibitem[\protect\citeauthoryear{{Waters}, {Di Matteo}, {Feng}, {Wilkins}  \&
  {Croft}}{{Waters} et~al.}{2016}]{Waters2016}
{Waters} D.,  {Di Matteo} T.,  {Feng} Y.,  {Wilkins} S.~M.,   {Croft} R. A.~C.,
   2016, \mn@doi [\mnras] {10.1093/mnras/stw2000}, \href
  {https://ui.adsabs.harvard.edu/abs/2016MNRAS.463.3520W} {463, 3520}

\bibitem[\protect\citeauthoryear{{White} et~al.,}{{White}
  et~al.}{2012}]{White2012}
{White} M.,  et~al., 2012, \mn@doi [\mnras] {10.1111/j.1365-2966.2012.21251.x},
  \href {https://ui.adsabs.harvard.edu/abs/2012MNRAS.424..933W} {424, 933}

\bibitem[\protect\citeauthoryear{{Williams} et~al.,}{{Williams}
  et~al.}{2018}]{Williams2018}
{Williams} C.~C.,  et~al., 2018, \mn@doi [\apjs] {10.3847/1538-4365/aabcbb},
  \href {https://ui.adsabs.harvard.edu/abs/2018ApJS..236...33W} {236, 33}

\bibitem[\protect\citeauthoryear{{Willott} et~al.,}{{Willott}
  et~al.}{2010}]{Willott2010}
{Willott} C.~J.,  et~al., 2010, \mn@doi [\aj] {10.1088/0004-6256/139/3/906},
  \href {https://ui.adsabs.harvard.edu/abs/2010AJ....139..906W} {139, 906}

\bibitem[\protect\citeauthoryear{{Yue} et~al.,}{{Yue} et~al.}{2018}]{Yue2018}
{Yue} B.,  et~al., 2018, \mn@doi [\apj] {10.3847/1538-4357/aae77f}, \href
  {https://ui.adsabs.harvard.edu/abs/2018ApJ...868..115Y} {868, 115}

\makeatother
\end{thebibliography}




\appendix

\section{Expected number counts}

\begin{figure*}
    \begin{center}
      \includegraphics[width=170mm]{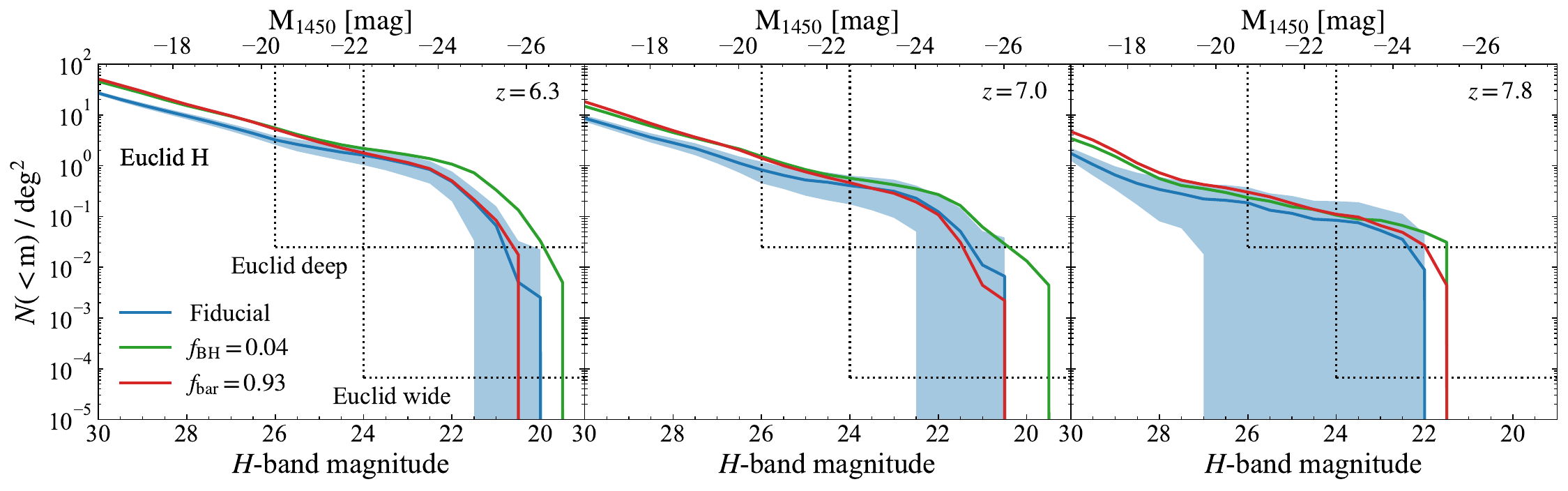}
    \end{center}
  \caption{Expected number of AGN brighter than a given apparent $H$-band magnitude per square degree, showing the fiducial galaxy model (blue), model with $f_{\mathrm{BH}}=0.04$ (green), and model with $f_{\mathrm{bar}}=0.93$ (red). Shaded regions show the total field-to-field variance of the fiducial model when assuming a $40~\mathrm{deg}^2$ survey area, corresponding to the Euclid deep survey. 
  Only the variance for the fiducial models is shown in order to make the figure easy to see.
  The amount of variances for the other two models are similar to the fiducial model. Dotted lines correspond to the magnitude limit and survey area of Euclid deep ($145 \mathrm{nJy}$, $40~\mathrm{deg}^2$) and Euclid wide ($912 \mathrm{nJy}$, $15000~\mathrm{deg}^2$). 
  }
  \label{fig:cumLF_Euclid}
\end{figure*}

In Fig.~\ref{fig:cumLF_Euclid}, we show the expected angular number density of bright AGN for these three models. We predict that 120--240 (16--80) AGN -- depending on the model -- with rest-frame UV absolute magnitude brighter than $-20$ ($-20.5$) will be observed at $z=6.3$ (7) in the Euclid H-band deep survey. For the Euclid wide survey, we predict that 15000--45000 (3000--15000) AGN with UV magnitude brighter than $-22$ ($-22.5$) will be detectable at $z=6.3$ (7).

\begin{figure*}
    \begin{center}
      \includegraphics[width=170mm]{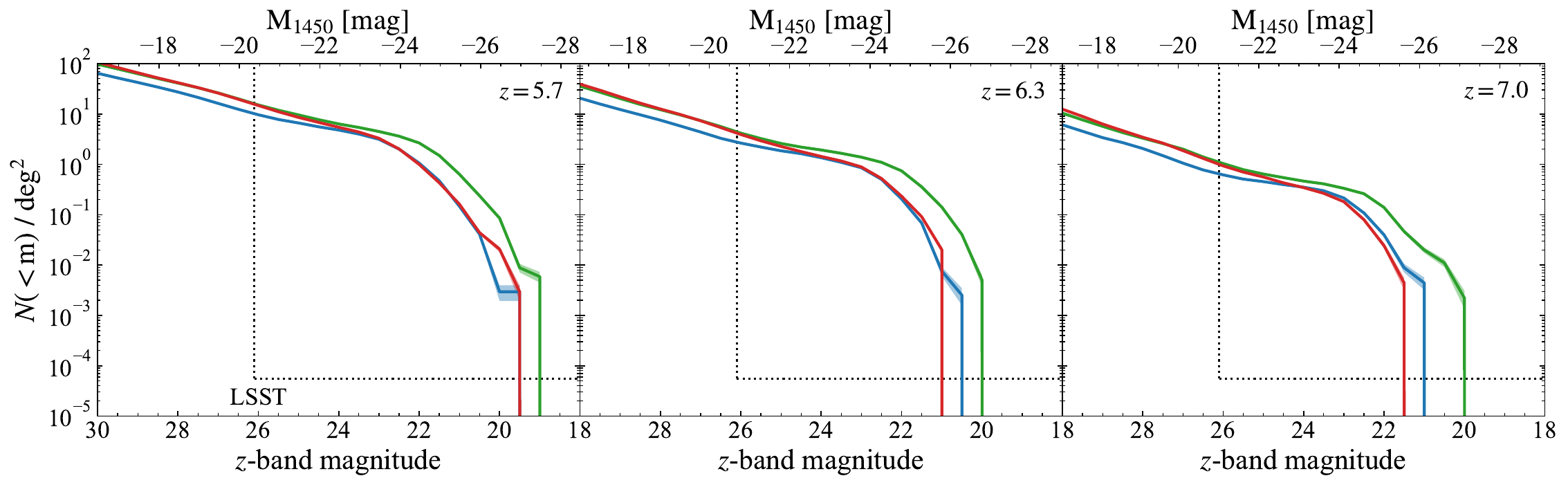}
    \end{center}
  \caption{Same as for Fig.~\ref{fig:cumLF_Euclid}, but in the LSST z-band. Shaded regions show the total field-to-field variance when assuming a $18000~\mathrm{deg}^2$ survey area, corresponding to the LSST survey. Dotted lines correspond to the magnitude limit and survey area of the LSST survey ($26.1~\mathrm{mag}$ in the z-band, $18000~\mathrm{deg}^2$).
  }
  \label{fig:cumLF_LSST}
\end{figure*}

In Fig.~\ref{fig:cumLF_LSST}, we also show the expected angular number density of bright AGN for these three models in the case of the LSST. We predict that 54000--72000 (10800--18000) AGN -- depending on the model -- with rest-frame UV absolute magnitude brighter than $-20.5$ ($-21.0$) will be observed at $z=6.3$ (7) in the LSST survey.

\section{Data base release}

All Uchuu-$\nu^2$GC data used for this paper is publicly available in our Skies \& Universes website. While we refer to Section~\ref{subsec:model} (and references therein) for a description of the model, we list in Table~\ref{tab:database} a subset of the galaxy properties included in the Uchuu-$\nu^2$GC public catalog. We further encourage the reader to visit the database for the additional documentation provided there, as the list of galaxy properties is not limited to what is shown in Table~\ref{tab:database}.

\begin{savenotes}
\begin{table*}
  \caption{Set of galaxy properties in the Uchuu-$\nu^2$GC catalog. Note that $x,y,z,v_x,v_y,v_z$, HaloMass, and Vmax for orphan galaxies have been set to those for the host halo in Uchuu-$\nu^2$GC. Also note that additional properties to those listed here have been included in the publicly available Uchuu-$\nu^2$GC data; refer to the database website for more information.
  }
\label{tab:database}
\begin{center}
\begin{tabular}{lll}
\hline
database name			& unit 			& description\\
\hline
\hline
HostHaloID			& n/a				& Pointer to dark matter halo in which galaxy resides; identical to id in\\
					&				&  the \textsc{rockstar} halo catalog, not applicable for orphan galaxies\\
MainHaloID			& n/a				& Pointer to dark matter halo in which galaxy orbits\\
GalaxyType			& n/a				& 0 = central galaxy\\
					&				& 1 = satellite galaxy\\
					&				& 2 = orphan galaxy\\
X					& comoving \hMpc	& $x$-position of galaxy\\
Y					& comoving \hMpc	& $y$-position of galaxy\\
Z					& comoving \hMpc	& $z$-position of galaxy\\
Vx					& peculiar \kms		& $v_x$-velocity of galaxy \\
Vy					& peculiar \kms		& $v_y$-velocity of galaxy	 \\
Vz					& peculiar \kms		& $v_z$-velocity of galaxy\\
MstarBulge			& \hMsun			& Stellar mass of bulge component of galaxy\\
MstarDisk				& \hMsun			& Stellar mass of disc component of galaxy\\
McoldBulge			& \hMsun			& Cold gas mass of bulge component of galaxy\\
McoldDisk				& \hMsun			& Cold gas mass of disc component of galaxy\\
Mhot					& \hMsun			& Total hot gas mass in galaxy\\
Mbh					& \hMsun			& Mass of central black hole\\
SFR					& \hMsun~\invGyr		& Star formation rate\\
MeanAgeStars			& Gyr			& Rest frame V-band luminosity weighted stellar age\\
HaloMass				& \hMsun			& $M_{\rm vir}$ of galaxy's dark matter halo\\
Vmax				& km/s			& Peak circular rotation velocity of galaxy's dark matter halo\\
Concentration		& n/a				& Concentration of galaxy's dark matter halo, -1 for orphan galaxies\\
SpinParameter			& n/a				& Spin parameter $\lambda$ of galaxy's dark matter halo, -1 for orphan galaxies\\
ZstarBulge		& n/a			& Mass-weighted mean stellar metallicity of bulge, -99 for no bulge stars\\
ZstarDisk			& n/a			& Mass-weighted mean stellar metallicity of disc, -99 for no disc stars\\
MZgasDisk			& ${\rm{M_{\odot}}}$			& Mass of metals in gas component of disc\\
LumAgnBol			& erg/s			& AGN bolometric luminosity\\
LumAgnXray			& erg/s			& AGN hard X-ray (2-10keV) luminosity\\
MagAgnUV			& n/a			& AGN UV (1450~\AA) magnitude, 128 for no AGN activity\\
\hline
\underline{Rest-frame magnitudes:}\\
MagStar(d)\_SDSSu		& n/a				& Dust uncorrected (corrected) absolute magnitude in SDSS u band\\
MagStar(d)\_SDSSg		& n/a				& Dust uncorrected (corrected) absolute magnitude in SDSS g band\\
MagStar(d)\_SDSSr			& n/a				& Dust uncorrected (corrected) absolute magnitude in SDSS r band\\
MagStar(d)\_SDSSi			& n/a				& Dust uncorrected (corrected) absolute magnitude in SDSS i band\\
MagStar(d)\_SDSSz			& n/a				& Dust uncorrected (corrected) absolute magnitude in SDSS z band\\
MagStar(d)\_2MASSJ			& n/a				& Dust uncorrected (corrected) absolute magnitude in 2MASS J band\\
MagStar(d)\_2MASSH			& n/a				& Dust uncorrected (corrected) absolute magnitude in 2MASS H band\\
MagStar(d)\_2MASSK			& n/a				& Dust uncorrected (corrected) absolute magnitude in 2MASS Ks band\\
MagStar(d)\_GALEXFUV			& n/a				& Dust uncorrected (corrected) absolute magnitude in GALEX FUV band\\
MagStar(d)\_GALEXNUV			& n/a				& Dust uncorrected (corrected) absolute magnitude in GALEX NUV band\\
\hline
\underline{Observer-frame magnitudes:}\\
AppMagStar(d)\_SDSSu		& n/a				& Dust uncorrected (corrected) apparent magnitude in SDSS u band\\
AppMagStar(d)\_SDSSg		& n/a				& Dust uncorrected (corrected) apparent magnitude in SDSS g band\\
AppMagStar(d)\_SDSSr			& n/a				& Dust uncorrected (corrected) apparent magnitude in SDSS r band\\
AppMagStar(d)\_SDSSi			& n/a				& Dust uncorrected (corrected) apparent magnitude in SDSS i band\\
AppMagStar(d)\_SDSSz			& n/a				& Dust uncorrected (corrected) apparent magnitude in SDSS z band\\

AppMagStar(d)\_HSCg		& n/a				& Dust uncorrected (corrected) apparent magnitude in HSC g band\\
AppMagStar(d)\_HSCr		& n/a				& Dust uncorrected (corrected) apparent magnitude in HSC r band\\
AppMagStar(d)\_HSCi		& n/a				& Dust uncorrected (corrected) apparent magnitude in HSC i band\\
AppMagStar(d)\_HSCz		& n/a				& Dust uncorrected (corrected) apparent magnitude in HSC z band\\
AppMagStar(d)\_HSCy		& n/a				& Dust uncorrected (corrected) apparent magnitude in HSC y band\\

\hline
\end{tabular}
\end{center}
\end{table*}
\end{savenotes}




\bsp	
\label{lastpage}
\end{document}